\documentclass[11pt]{article}
\usepackage[cp1251]{inputenc}
\usepackage[english]{babel}   % load Babel setup for English
\usepackage{amsfonts}
\usepackage{graphics}
\usepackage{amssymb}
\usepackage{amsmath}
\usepackage{epsfig}
\begin{document}
\sloppy

\begin{center}
\Large{\textbf{Influence of Multiple Scattering on High-energy
Deuteron Quasi-optical Birefringence Effect}}
\end{center}

\begin{center}
{Baryshevsky V.G.,  Shyrvel  A.R.}
\end{center}

\begin{center}
\textit{{Research Institute for Nuclear Problems, Minsk, Belarus}}
\end{center}

\bigskip

\section*{Introduction}

A quasi-optical macroscopic quantum phenomenon of birefringence, i.e., the effect of particle spin rotation (oscillation)
and spin dichroism accompanying the passage of high-energy particles with spin $S \ge 1$ through isotropic homogeneous matter
was first described in [1, 2].
This effect is analogous to the effect known in optics as the birefringence of light in
an optically anisotropic medium associated with the dependence of the index of refraction on
the state of light polarization, i.e., on the photon spin state.
In contrast to light, whose wavelength is much longer than the distance between the atoms of matter, the de Broglie
wavelength of a fast deuteron is much smaller than this distance. Nevertheless, according to [1, 2], in this case
 one can also introduce the spin-dependent index of refraction for particles.
The deuteron birefringence effect occurs even in a homogeneous isotropic medium and is due to the intrinsic anisotropy
inherent in particles with spin $S \ge 1$. Moreover, according to [1, 2],
the magnitude of this effect even increases with growing particle energy, i.e., with diminishing wavelength.
It was shown in [1, 2]
that the spin dichroism described by the imaginary part of the index of refraction causes tensor polarization of
the initially unpolarized deuteron beam. Recently, this effect was experimentally revealed in the energy regions $5\div20$ MeV [3, 4]
and 3 GeV  [5].
According to [4, 5], %\cite{4,5}
the appearance of tensor polarization due to spin dichroism in
particles with spin $S \ge 1$ can be used for designing the source
of deuterons, which possess tensor polarization. According to [6,
7, 8], the birefringence effect should also be taken into account
in experiments to search for deuteron EDM.

 In theoretical description of the birefringence phenomenon, it should be
taken into account that the amplitude of forward scattering of a deuteron by a nucleus, which  determines the index of refraction, includes
the contribution due to Coulomb interaction alongside with that due to strong interaction. The Coulomb-nuclear interference correction to
the rotation angle of the polarization vector and spin dichroism of the deuteron was computed
in [2, 9]. % \cite{2,7}.
According to [9], the Coulomb-nuclear interference enables
explanation of the change of sign of spin dichroism with changing
deuteron energy in the range of $5\div 20$ MeV.

It should be noted that the interaction of deuterons and matter is accompanied not only by coherent scattering,
which leads to the formation of a coherent wave, but also by incoherent scattering resulting in single and multiple
scattering of particles in the target.
In a kinetic equation for the density matrix, the characteristics of elastically scattered particles are described by the terms proportional to the squared absolute value of the total scattering amplitude (this amplitude is the sum of the Coulomb amplitude and the nuclear scattering amplitude modified by the Coulomb interaction) and as a result, in this case the terms describing the interference of the Coulomb and nuclear interactions should also be considered.
In this connection, [10, 11] give the analysis of the correction
due to the stated interference terms to the rotation of the
polarization vector of protons passing through polarized matter
for such target thicknesses and scattering angles at which the
particle angular distribution is determined by the multiple
Coulomb scattering and the single nuclear scattering in matter.
It was shown in [11] that in the expression for the spin rotation
angle of a high energy proton beam, the contribution of the
Coulomb-nuclear interference to the amplitude of nuclear
scattering at zero angle is compensated by the correction
from the incoherent Coulomb-nuclear scattering. This statement was also formulated for scattering of particles with spin 1 [12].%

In the present paper is shown that within the domain of applicability of the first Born approximation, for the Coulomb
amplitude the contributions to the rotation angle of the deuteron polarization vector coming from the Coulomb-nuclear
interference are compensated only in the case when the detector provides a  maximum coverage within  a $4\pi$ solid angle geometry.
If the detector registers the particles moving within a certain
solid angle $\Delta \Omega\ll 4\pi$ with respect to the initial
direction of the beam incidence, as it does in a real experiment,
the interference terms stated above are not compensated.
The magnitude of the contribution of the Coulomb-nuclear
interference depends appreciably on the specific geometry of the
experiment and is quite observable in the experiments on measuring
the spin rotation angle of particles passing through the target.
Multiple scattering of particles in the target leads to the fact
that the magnitude of the contribution of the Coulomb-nuclear
interference to the rotation angle of the beam polarization vector
depends not only on the angle $\Delta\Omega$ of the detector, but also on the
mean square angle of multiple scattering.
It is shown that if after the beam has passed the target, the
magnitude of the mean square angle of multiple scattering  is of
the order of magnitude or greater than the squared diffraction
angle of nuclear scattering, the Coulomb-nuclear interference can
be neglected for any registered angle $\Delta\Omega$ of the detector. Otherwise, the
contribution of the total Coulomb-nuclear interference to the
rotation angle should be taken into account.

\section{Kinetic equation for the spin density matrix}

Let us consider the process of particle (beam of particles)
passage through a target consisting of $N_{t}$ particles
interacting with one another by means of a certain potential $U$.

Suppose that an incident particle has a rest mass $m$  and spin
$S_{d}$. The target consists of $N_{t}$ bound particles with mass
$M$ and spin $s$.

The Hamiltonian of the scattering system is written in the form:
$$
\begin{array}{l}
\displaystyle H_{t}=\sum_{\alpha=1}^{N_{t}}K_{\alpha}+U,
\end{array}\eqno(1.1)
$$
 where $K_{\alpha}$ is the kinetic energy operator of particle
$\alpha$, $U$ is the interaction energy of  $N_{t}$ particles of
the scatterer.

The solution of the stationary Schr\"{o}dinger  equation
$$
\begin{array}{l}
\displaystyle
H_{t}\Psi_{\gamma}=W_{\gamma}\Psi_{\gamma}
\end{array}\eqno(1.2)
$$
determines the possible values of the energy $W_{\gamma}$ of the system  and the corresponding set of wave functions.
$$
\begin{array}{l}
\displaystyle
\Psi_{\gamma}=\Psi_{\gamma}(\vec{R}_{1},s_{1},\ldots,\vec{R}_{N},s_{N}),
\end{array}\eqno(1.3)
$$
where $(\vec{R}_{\alpha},s_{\alpha})$ are the spatial and spin coordinates of particle
$\alpha$  in the scatterer.

The operator of interaction between the incident particle and the
scatterer is defined in terms of $V$: $\displaystyle
V=\sum_{\alpha=1}^{N_{t}}V_{\alpha}$.
Here $V_{\alpha}$ describes the interaction between the beam particle  and the target particle $\alpha$.
The Hamiltonian of the whole system has a form:
$$
\begin{array}{l}
\displaystyle H=K_{d}+H_{t}+V,
\end{array}\eqno(1.4)
$$
where $K_{d}$ is the kinetic energy operator of the beam particle $d$.

To describe the process of particle $d$ transmission through the target, find the density
matrix $\hat{\rho}(t)$ of the system "incident particle + target". This density matrix satisfies the quantum Liouville equation:
$$
\begin{array}{l}
\displaystyle i\frac{\partial\hat{\rho}}{\partial
t}=[\hat{H},\hat{\rho}(t)].
\end{array}\eqno(1.5)
$$
The solution of this equation can formally be written using the evolution operator $\hat{U}(t,t_{0})$
$$
\begin{array}{l}
\displaystyle
\hat{\rho}(t)=\hat{U}(t,t_{0})\hat{\rho}(t_{0})\hat{U}^{\scriptscriptstyle+}(t,t_{0}),
\end{array}\eqno(1.6)
$$
which is related to the explicitly time-independent Hamiltonian of
the system as
$\displaystyle\hat{U}(t,t_{0})=e^{-\frac{i}{\hbar}\hat{H}(t-t_{0})}$.
The time moments $t_{0}$ and $t$ correspond to the state of the
system before and after scattering of the incident particle by the
$\alpha$-th scatterer.

Let us consider the target as a thermostat ($N_{t}\gg1$). Then the
statistical operator $\hat{\rho}$ of the system can be represented
as a direct product
$\hat{\rho}=\hat{\rho}_{d}\otimes\hat{\rho}_{t}$, where
$\hat{\rho}_{d}$ is the density matrix of the incident particle,
$\hat{\rho}_{t}$ is the equilibrium density matrix of the medium.
The spin density matrix $\hat{\rho}_d$ includes the elements
diagonal and nondiagonal with respect to the momenta $\vec{k}$ of
the incident particle:
$\hat{\rho}_d=\hat{\rho}_d(\vec{k},\vec{k}^{\prime})$. However,
the nondiagonal elements oscillate fast and after several
collisions with target nuclei, in order to describe the process of
multiple scattering, one can assume that the density matrix
$\hat{\rho}_d(\vec{k},\vec{k}^{\prime})$ is diagonal [10, 13, 14,
15], i.e.,
$\hat{\rho}_d(\vec{k},\vec{k}^{\prime})=\delta_{\vec{k},\vec{k}^{\prime}}\hat{\rho}_d(\vec{k})$.

The  time interval $\Delta t$ during which the density matrix is
diagonalized satisfies the inequality $\Delta t\gg R/\bar{v}$,
where $R$ is the radius of action of the forces, $\bar{v}$ is the
particle mean velocity in matter. This means that the evolution
operator $U(t,t_{0})$ can be replaced by the Heisenberg's
$S$-matrix
$\displaystyle\equiv\lim_{{t\rightarrow\infty},{t_{0}\rightarrow-\,\infty}}U(t,t_{0})$,
which relates the asymptotic states of the system before
scattering to those after scattering [10, 13, 14, 15]. The matrix
elements of the $S$-matrix for a scattering system consisting of
$N_{t}$ particles are defined as follows:

$$
\begin{array}{l}
\displaystyle
S_{ba}=\delta_{ba}-i(2\pi)\delta(E_{b}-E_{a})\sum_{\alpha=1}^{N_{t}}\left(\mathcal{T}_{\alpha}\right)_{ba},
\end{array}\eqno(1.7)
$$
where $E_{a}$  and  $E_{b}$ are the total energies of the system
before and after scattering, respectively;
$E_a=\varepsilon_k+W_{\gamma}$,
$E_b=\varepsilon_{k^{\prime}}+W_{\gamma^{\prime}}$,
$\varepsilon_{{k}}$ and $\varepsilon_{{k}^{\prime}}$ are the
energies of the incident particle before and after the collision.
In the momentum approximation, choose as $\mathcal{T}_{\alpha}$
the scattering matrix of particle $d$ interacting with a free
particle $\alpha$ [16].

The limits $t_{0}\rightarrow-\,\infty$, $t\rightarrow\infty$ are understood as the
times when the incident particle is situated at distances larger than $R$, but the
interval $\Delta t=t-t_{0}$ is small as compared to $l/\bar{v}$ ($l$ is the mean free path of a particle in matter).

Denote the density matrix of the system at time $t=t_{0}+\Delta t$
by $\hat{\rho}'$, accordingly before scattering by the the
$\alpha$-th particle of the target, the density matrix will read
$\hat{\rho}$. Equation (1.6) can be rewritten for the density
matrix of the incident particle, using the $S$-matrix  defined in
equation (1.7):

$$
\begin{array}{l}
\displaystyle \hat{\rho}\,'_{d}=\texttt{Sp}_{t}S\hat{\rho}S^{+},
\end{array}\eqno(1.8)
$$
where $\texttt{Sp}_t$ means  taking the trace over the states of
the target.

Write equality (1.8) for the diagonal momentum-space elements of
the particle density matrix:

$$
\begin{array}{l}
\displaystyle
\hat{\rho}_{d}'(\vec{k})=\hat{\rho}_{d}(\vec{k})-i(2\pi)\frac{\Delta
t}{2\pi}\texttt{Sp}_{t}\sum_{\alpha=1}^{N_T}
\langle\vec{k},\gamma|\hat{\mathcal{T}}_{\alpha}|\vec{k},\gamma\rangle\hat{\rho}(\vec{k},\gamma)+
i(2\pi)\frac{\Delta
t}{2\pi}\texttt{Sp}_{t}\sum_{\alpha=1}^{N_T}\hat{\rho}(\vec{k},\gamma)
\langle\vec{k},\gamma|\hat{\mathcal{T}}_{\alpha}^{\scriptscriptstyle+}|\vec{k},\gamma\rangle+\\[-10pt]
{}\\
\displaystyle+(2\pi)^{2}\frac{\Delta
t}{2\pi}\texttt{Sp}_{t}\sum_{\alpha,\beta=1}^{N_T}\sum_{\vec{k}'\gamma'}
\langle\vec{k},\gamma|\hat{\mathcal{T}}_{\alpha}|\vec{k}',\gamma'\rangle\hat{\rho}(\vec{k}',\gamma')
\delta(\varepsilon_{k'}-\varepsilon_{k}+W_{\gamma'}-W_{\gamma})
\langle\vec{k}',\gamma'|\hat{\mathcal{T}}_{\beta}^{\scriptscriptstyle+}|\vec{k},\gamma\rangle.
\end{array}\eqno(1.9)
$$

In the momentum space, the matrix elements for the scattering
matrix operator $\hat{\mathcal{T}}$ have a form [17]:

$$
\begin{array}{l}
\displaystyle
\langle\vec{k}',\vec{P}_{\alpha}'|\hat{\mathcal{T}}_{\alpha}|\vec{k},\vec{P}_{\alpha}\rangle=(2\pi)^{3}
\delta(\vec{k}'+\vec{P}_{\alpha}'-\vec{k}-\vec{P}_{\alpha})
\langle\vec{k}',\vec{P}_{\alpha}'|\hat{\texttt{T}}_{\alpha}|\vec{k},\vec{P}_{\alpha}\rangle,
\end{array}\eqno(1.10)
$$
$\vec{P}_{\alpha}$ denotes the momenta of the  $\alpha$-th
scatterer, $\hat{\texttt{T}}_{\alpha}$ is the matrix of scattering
by the momentum shell.
Recall that  $\hat{\mathcal{T}}_{\alpha}$ and
$\hat{\texttt{T}}_{\alpha}$ remain the operators with respect to spin variables.

The states with a definite value of the momentum $|\vec{k}\rangle$
are normalized according to equality
$\displaystyle\langle\vec{k}'|\vec{k}\rangle=\frac{(2\pi)^{3}}{V}\,\delta(\vec{k}'-\vec{k})$,
where $V$ is the normalization volume. Thus, substitution of
summation over all $\vec{k}$ by integration is made as follows
$\displaystyle\sum_{\vec{k}}\rightarrow
\frac{V}{(2\pi)^{3}}d^{3}\vec{k}$; the Kronecker symbol and the
Dirac function  $\delta$ are related as
$\displaystyle\delta_{\vec{k}\vec{k}'}\rightarrow\frac{(2\pi)^{3}}{V}\delta(\vec{k}-\vec{k}')$.

Scattering processes that make the contribution to (1.9) can be
both elastic and inelastic. Elastic scattering occurs in the
absence of any excitation in the scatterer. Inelastic scattering
accompanied by a small excitation in the scatterer is called
"quasi-elastic" \, [16]. For such scattering the momentum
$\vec{q}$ transferred to the $\alpha$-th scatterer is
$q\ll\sqrt{2M K_{\alpha}}$, $q\ll\sqrt{2M U_{\alpha}}$,
$\vec{q}=\vec{k}'-\vec{k}$; $\vec{k}$ is the momentum of the
incident particle before the collision with the scatterer,
$\vec{k}'$ is the momentum of the beam particle after the
collision.

When the momentum transferred to the scatterer $\alpha$ exceeds
the momentum of the target particle in the initial bound state
$q\gg\sqrt{2M K_{\alpha}}$, $q\gg\sqrt{2M U_{\alpha}}$, then the
so-called "quasi-free"  approximation  can be used. In this
approximation the excitation energy of the system in one
collision is exactly equal to the recoil energy $\vec{q}\,^2/2M$
of a free particle of the target. Note that in summation over
different nuclei of the target, sums of the form given below
appear in the last term in (1.10)

$$
\begin{array}{l}
\displaystyle \sum_{\alpha,\beta=1}^{N_{t}}\int
d^{3}\vec{R}_{1}...d^{3}\vec{R}_{N}
e^{-i\vec{q}(\vec{R}_{\alpha}-\vec{R}_{\beta})}\hat{\rho}_{t}(\vec{R}_{1},...\vec{R}_{N})=N_{t}\hat{\rho}_{t}+\\[-10pt]
{}\\
\displaystyle+\sum_{\alpha\neq\beta}^{N_t}\int
d^{3}\vec{R}_{1}...d^{3}\vec{R}_{N}
e^{-i\vec{q}(\vec{R}_{\alpha}-\vec{R}_{\beta})}\hat{\rho}_{t}(\vec{R}_{1},...\vec{R}_{N}),
\end{array}\eqno(1.11)
$$
where $\hat{\rho}_t$ is the spin density matrix of the target
nucleus. In deriving equation (1.11) it is assumed that the
positions of target nuclei and their spin states are uncorrelated.

 Upon averaging over the spin states of nuclei in the target,
the second term in (1.11) can be expressed in terms of the
so-called particle pair distribution function [16], and vanishes
when the transferred momentum $\vec{q}$ exceeds the magnitude
inverse to the correlation radius $r$, i.e.,  $\displaystyle q\gg
r^{-1}$ [16]. For a non-crystalline target, the magnitude of the
correlation radius $r$ is of the order of the distance between
nuclei. Consequently, the second term can only contribute to the
kinetic equation at very small scattering angles
$\vartheta_{sc}\lesssim{1}/{k r}$, so it will be neglected in
further consideration.

The resulting expression for the density matrix can be obtained
from (1.9) for
$\displaystyle\frac{\hat{\rho}^{\prime}_d-\hat{\rho}_d}{\Delta t}$
in the form of the following integro-differential equation:
$$
\begin{array}{l} \displaystyle
\frac{d\hat{\rho}_{d}(\vec{k},t)}{dt}=- iVN_{t}
\texttt{Sp}_{t}\left(\hat{\texttt{T}}(\vec{k},\vec{k})
\hat{\rho}(\vec{k},t)-\hat{\rho}(\vec{k},t)\hat{\texttt{T}}^{\scriptscriptstyle+}(\vec{k},\vec{k})\right)
+\\[-10pt]
{}\\
\displaystyle+\frac{V^{3}}{(2\pi)^{2}}N_{t}\texttt{Sp}_{t}\int
d^{3}\vec{k}'
\delta\left(\varepsilon_{k}-\varepsilon_{k'}-\frac{\vec{q}^{2}}{2M}\right)
\hat{\texttt{T}}(\vec{k},0;\vec{k}',-\vec{q})\hat{\rho}(\vec{k}',t)
\hat{\texttt{T}}^{\scriptscriptstyle+}(\vec{k}',-\vec{q};\vec{k},0).
\end{array}\eqno(1.12)
$$
$\hat{\rho}(\vec{k})$ denotes the following dependnce:
$\displaystyle\hat{\rho}(\vec{k})=\hat{\rho}_{d}(\vec{k};\vec{S}_{d})\otimes\hat{\rho}_{t}(\vec{S}_{t})$,
where $\hat{\rho}_t(\vec{S}_{t})$ is the spin density matrix of
the target nucleus.

Let us introduce the scattering amplitude $\hat{F}$ with matrix
elements equal to [17]:

$$
\begin{array}{l} \displaystyle
\hat{F}(\vec{k},\vec{k}')=-\frac{M_{r}}{2\pi}V^{2}\hat{\texttt{T}}(\vec{k},0;\vec{k}',-\vec{q}),
\end{array}\eqno(1.13)
$$
where $\displaystyle M_{r}=\frac{mM}{m+M}$ is the reduced mass.

Then equation (1.12) can be transformed into a form:

$$
\begin{array}{l} \displaystyle
\frac{d\hat{\rho}_{d}(\vec{k},t)}{dt}=\frac{2\pi i}{M_{r}}N
\texttt{Sp}_{t}\left(\hat{F}(\vec{k},\vec{k})
\hat{\rho}(\vec{k},t)-\hat{\rho}(\vec{k},t)\hat{F}^{\scriptscriptstyle+}(\vec{k},\vec{k})\right)
+\\[-10pt]
{}\\
\displaystyle+N\texttt{Sp}_{t}\int d\Omega_{\vec{k}'}
\frac{k'^{2}}{M_{r}^{2}\left(\frac{k'}{m}-\frac{(\vec{k}-\vec{k}')\vec{n}'}{M}\right)}
\hat{F}(\vec{k},\vec{k}')\hat{\rho}(\vec{k}',t)
\hat{F}^{\scriptscriptstyle+}(\vec{k}',\vec{k}),
\end{array}\eqno(1.14)
$$
where $N$  denotes the number of  particles in the target per unit
volume, $\vec{n}'$ is the unit vector in the direction of the
momentum $\vec{k}'$. The absolute value of vector $\vec{k}'$ is
determined from the equation:
$$
\begin{array}{l} \displaystyle
\varepsilon_{k}=\varepsilon_{k'}+\frac{(\vec{k}-\vec{k}')^{2}}{2M}.
\end{array}\eqno(1.15)
$$
Note that when the condition $m\geq M$ for the masses of the
incident particles is fulfilled, the denominator in the integrand
of equation (1.14) vanishes for the value of the scattering angle
of the incident particle equal to
$\displaystyle\cos\theta=\frac{\sqrt{m^{2}-M^{2}}}{m}$, with the
absolute value of vector $\vec{k}'$ being equal to $\displaystyle
k\sqrt{\frac{m-M}{m+M}}$.

Equation (1.14) simplifies when a particle (proton, deuteron,
antiproton) passes through a target with nuclei whose mass is much
larger than the mass of the incoming particle. In this case we can
neglect the effect of the energy loss of the incident particle
through scattering. So we can neglect the recoil energy
$\displaystyle{\vec{q}^2}/{2M}$ in the $\delta$-function (1.12),
(1.15). As a result, one obtains a simple kinetic equation
describing the time and spin evolution of the incident particle as
it passes through the target [10]:

$$
\begin{array}{l} \displaystyle
\frac{d\hat{\rho}_d(\vec{k},t)}{dt}=\frac{2\pi N}{m}
\texttt{Sp}_{t}\left[\hat{F}(\vec{k},\vec{k})
\hat{\rho}_d(\vec{k},z)-\hat{\rho}_d(\vec{k},z)\hat{F}^{\scriptscriptstyle+}(\vec{k},\vec{k})\right]
+N\frac{k}{m}\texttt{Sp}_{t}\int d\Omega_{\vec{k}'}
\hat{F}(\vec{k},\vec{k}')\hat{\rho}(\vec{k}',t)
\hat{F}^{\scriptscriptstyle+}(\vec{k}',\vec{k}),
\end{array}\eqno(1.16)
$$
where $|\vec{k}|=|\vec{k}'|$.

The first term on the right-hand side of (1.16), which describes
refraction of particle in the target, can be represented as
follows:

$$
\hat {F}(0)\hat {\rho }(\vec {k},t) - \hat {\rho }(\vec {k},t)\hat
{F}^ + (0) = \left[ {\textstyle{1 \over 2}\left( {\hat {F}(0) +
\hat {F}^{{\scriptscriptstyle +}} (0)}\right),\hat {\rho }(\vec
{k},t)} \right] + \left\{ {\textstyle{1 \over 2}\left( {\hat
{F}(0) - \hat {F}^{\scriptscriptstyle +} (0)} \right),\hat {\rho
}(\vec {k},t)} \right\},\eqno(1.17)
$$

\noindent where $\left[ , \right]$ is the commutator, $\left\{\, ,
\right\}$ is the anticommutator.

The part proportional to the commutator leads to the rotation of
the polarization vector due to elastic coherent scattering (as a
result of the refraction effect [10]); the anticommutator
describes the reduction in the intensity and depolarization of the
beam which has passed through the target. The last term in (1.16)
determines the effect of incoherent scattering on the change of
$\hat{\rho}_d$ (in the general case, single and multiple
scattering).

As stated above, equation (1.16) is not applicable to the description of the process
of proton (deuteron) transmission through the target containing light nuclei (protons (deuterons)).
To describe multiple scattering in this case, a more general equation (1.14) should be solved.

Further we shall concern ourselves with the deuteron passage
through a carbon target, and so we shall use equation (1.16).

For the sake of concreteness, let us consider the process of
deuteron passage through the target with spinless nuclei. In this
case the density matrix $\hat {\rho}(\vec {k})$, as well as the
amplitude $\hat {F}(\vec {k},{\vec {k}}')$ contain only spin
variables of the scattered beam: $\hat {\rho }(\vec {k}) =
\hat{\rho}_d (\vec {k})$. For the amplitude $\hat {F}(\vec
{k},{\vec {k}}')$, we shall introduce the notation $\hat
{F}(\vec{k},{\vec {k}}') = \hat {f}(\vec {k},{\vec {k}}')$, where
$\hat {f}(\vec {k},{\vec {k}}') \equiv \hat {f}(\vec {k},{\vec
{k}}';\hat{\vec {S}}_{d})$.

As a result, equation (1.16) can be written as follows:

$$
\begin{array}{l}
 \displaystyle\frac{d\hat {\rho }_d }{dz} = \frac{\pi i}{k}N\left[
{(\hat {f}(\vec {k},\vec {k}) + \hat {f}^{\scriptscriptstyle
+}(\vec {k},\vec {k})),\hat {\rho }_{d }(\vec {k})} \right] +
\frac{\pi i}{k}N\left\{ {(\hat {f}(\vec {k},\vec {k}) - \hat
{f}^{\scriptscriptstyle +}(\vec {k},\vec {k})), \hat {\rho
}_d (\vec {k})} \right\} + \\[-10pt]
{}\\
 \displaystyle+ N\int {d\Omega _{\vec {k}}' \hat {f}(\vec {k},{\vec {k}}') \hat
{\rho }_d ({\vec {k}}')\hat {f}^{\scriptscriptstyle
+}(\vec {k}',{\vec {k}})}, \\
 \end{array}\eqno(1.18)
$$
where $z=vt$ ($v$ is the particle velocity) is the distance
traveled by the incident particle in matter. Hereinafter, the
subscript $d$ of the density matrix will be dropped.

\section{Scattering Amplitude}

For particles with spin 1 (deuterons), the amplitude $\hat
{f}(\vec {k},{\vec {k}}')$  can be expressed in terms of the deuteron spin operator $\hat {\vec {S}}$,
quadrupolarization tensor $\hat{Q}_{ik} $ and the combination of vectors $\vec {k}$ and ${\vec {k}}'$:

$$ \hat {f}(\vec {k},{\vec {k}}') = A{\hat {I}} + B(\hat {\vec
{S}}\vec {\nu }) + C_1 \hat {Q}_{ik} \mu _i \mu _k + C_2 \hat
{Q}_{ik} \mu _{1i} \mu _{1k},\eqno(2.1)
$$

\noindent where $A$, $B$, $C_1 $ and $C_2 $ are the parameters
depending on $\theta $, $\vec {\nu } = [\vec {k}\times {\vec
{k}}'] / \vert [\vec {k}\times {\vec {k}}']\vert $, $\vec {\mu } =
(\vec {k} - {\vec {k}}') / \vert \vec {k} - {\vec {k}}'\vert $, $
\vec {\mu }_1 = (\vec {k} + {\vec {k}}') / \vert \vec {k} + {\vec
{k}}'\vert $, the components of the tensor  $\hat {Q}_{ik} $ are
defined as: $\hat {Q}_{ik} = \textstyle{3 \over 2}\left( {\hat
{S}_i \hat {S}_k + \hat {S}_k \hat {S}_i - \textstyle{4 \over
3}\delta _{ik} {\hat {I}}} \right)$, ${\hat {I}}$ is the $3\times
3$ identity matrix.

Write the explicit form of the zero-angle scattering amplitude $\hat {f}(\vec {k},\vec {k})$. In view of (2.1), we have:

$$ \hat {f}(\vec {k},\vec {k}) = f_0 (0) + f_1 (0)(\vec {S}\vec
{n})^2,\eqno(2.2)
$$
\noindent where $\vec {n} = \vec {k} / k$ is the unit vector in the direction  $\vec {k}$
and the following notations are introduced: $f_0 = A - 2C_1 - 2C_2 $, $f_1 = 3C_2 $.

In the general case $f_0 $ and $f_1 $ are the complex functions,
and according to the optical theorem, the imaginary parts  of $f_0
$ and $f_1 $ can be expressed in terms of the corresponding total
cross-sections:

$$ {\begin{array}{*{20}c}
\displaystyle {\mathrm{Im} f_0 (0) = \frac{k}{4\pi }\,\sigma
_{tot}^0 ,\,\,\,\,\,\,\,\,\mathrm{Im}f_1 (0) = \frac{k}{4\pi
}\left[ {\sigma _{tot}^{\pm 1} - \sigma _{tot}^0 } \right],}
\hfill \\
\end{array}}\eqno(2.3)
$$
where $\sigma _{tot}^0 $, $\sigma _{tot}^{\pm 1} $ are the total scattering cross-sections
for the initial spin state of the deuteron with a magnetic quantum number $M = 0$ and $M = \pm 1$,
respectively (the quantization axis $z$ is directed along $\vec {n})$.

The deuteron interacts with the target nuclei via nuclear and Coulomb interactions. The  amplitude $\hat {f}(\vec
{k},{\vec {k}}')$ of the deuteron scattering by a target nuclei in this case can be represented in the form:

$$ \hat {f}(\vec {k},{\vec {k}}') = \hat {f}_{\scriptstyle coul} (\vec {k},{\vec
{k}}') + \hat {f}_{nucl,coul} (\vec {k},{\vec {k}}'),\eqno(2.4)
$$
\noindent where $\hat {f}_{coul} $ is the amplitude of the Coulomb
scattering of the deuteron by a
 nucleus in the absence of nuclear interaction, $\hat {f}_{nucl,coul} $ is the amplitude of scattering of
 Coulomb-distorted waves by a nuclear potential.

The matrix elements of the amplitude $f_{ba}$ in the case of
scattering by a fixed center are related to the elements of the
operator $\mathcal{T}$ as [17]:

$$f_{ba} = - \frac{m}{2\pi}V\mathcal{T}_{ba}.\eqno(2.5)
$$

In the case of two interactions, the  matrix elements of the operator $\mathcal{T}$ are defined in a standard manner
$$
\mathcal{T}_{ba} =\langle\Phi_{b}|V_{coul}+V_{nucl}|\psi_{a}^{+}\rangle.\eqno(2.7)
$$
 $\Phi_{{a(b)}}$ describes the initial (final) state of the system "particle--nucleus" \, in the area,
 where the interaction is absent, the wave function $\psi_{a}^{+}$ satisfies the integral equation
 $\displaystyle\psi_{a}^{+}=\Phi_{a}+(\varepsilon_{a}-K_{d}+i\eta)^{-1}(V_{coul}+V_{nucl})\psi_{a}^{+}$,
 where $K_{d}$ is the  kinetic energy operator of the incident particle (deuteron).

Upon introducing a wave function $\varphi^{-}_{b}$, which
describes a converging wave in scattering by a Coulomb potential
alone and corresponds to a final state $\Phi_{b}$:
$\varphi_{b}^{-}=\Phi_{b}+(\varepsilon_{a}-K_{d}-i\eta)^{-1}V_{coul}\varphi_{b}^{-}$,
the matrix elements of operator $\mathcal{T}$ can be represented as a sum of two terms [18]:%

$$\mathcal{T}_{ba} = \mathcal{T}_{ba}^{coul} + \mathcal{T}_{ba}^{nucl,coul}
,\eqno(2.8)
$$

\noindent where the matrix elements $\mathcal{T}_{ba}^{coul} $
correspond to the Coulomb interaction
$\mathcal{T}_{ba}^{coul}=\left\langle{\Phi _b}
\right|V_{coul}\left| {\varphi _a^{ +} } \right\rangle$, the part
$\mathcal{T}_{ba}^{nucl,coul} \equiv \left\langle {\varphi _b^{-}}
\right|V_{nucl} \left| {\psi_a^{+} } \right\rangle $ determines
the amplitude of scattering by a nuclear potential of waves that
have been scattered by the potential $V_{coul} $.

{Operator }$\mathcal{T}_{nucl,coul}$  can be expressed in terms of
the operators of the Coulomb scattering and nuclear scattering as
the following infinite series:

$$
\begin{array}{l}
 \mathcal{T}_{nucl,coul}=\mathcal{T}_{nucl} + \mathcal{T}_{nucl} G_0
\mathcal{T}_{coul} + \mathcal{T}_{coul} G_0 \mathcal{T}_{nucl} + \\[-10pt]
{}\\
\displaystyle + \mathcal{T}_{coul} G_0 \mathcal{T}_{nucl} G_0
\mathcal{T}_{coul} + \mathcal{T}_{nucl} G_0 \mathcal{T}_{coul} G_0
\mathcal{T}_{nucl} + \mathcal{T}_{nucl} G_0 \mathcal{T}_{coul} G_0
\mathcal{T}_{nucl} G_0 \mathcal{T}_{coul} + \\[-10pt]
{}\\
+\mathcal{T}_{coul}
G_0 \mathcal{T}_{nucl} G_0 \mathcal{T}_{coul} G_0 \mathcal{T}_{nucl} + ..., \\
 \end{array}\eqno(2.9)
$$

\noindent where $G_0$ is the stationary Green's function. By
definition $\displaystyle G_0 = \frac{1}{\varepsilon_a - K_{d} +
i\eta }$.

In view of the definition (2.5), equation (2.9) can be written, using the corresponding scattering amplitudes:
$$
\begin{array}{l}
 \displaystyle\hat {f}_{nucl,coul} (\vec {k},{\vec {k}}') = \hat {f}_{nucl} (\vec
{k},{\vec {k}}') - \frac{1}{(2\pi )^2m}\int {\frac{\hat {f}_{nucl}
(\vec {k},{\vec {k}}'')\hat {f}_{coul} ({\vec {k}}'',{\vec
{k}}')}{\varepsilon_{k} -
\varepsilon_{k''} + i\eta }} d^3{\vec {k}}'' -  \\[-10pt]
{}\\
\displaystyle
 - \frac{1}{(2\pi )^2m}\int {\frac{\hat {f}_{coul} (\vec {k},{\vec
{k}}'')\hat {f}_{nucl} ({\vec {k}}'',{\vec {k}}')}{\varepsilon_{k}
- \varepsilon_{k''}
+ i\eta }} d^3{\vec {k}}'' +  \\[-10pt]
{}\\
\displaystyle+\frac{1}{(2\pi )^4m^2}\int\!\!\!\int {\frac{\hat
{f}_{nucl} (\vec {k},{\vec {k}}'')\hat {f}_{coul} ({\vec
{k}}'',{\vec {k}}''')\hat {f}_{nucl} ({\vec {k}}''',{\vec
{k}}')}{(\varepsilon_{k} - \varepsilon_{k''} + i\eta
)(\varepsilon_{k} - \varepsilon_{k'''} + i\eta
)}} d^3{\vec {k}}''d^3{\vec {k}}''' +  \\[-10pt]
{}\\
\displaystyle
 + \frac{1}{(2\pi )^4m^2}\int\!\!\!\int {\frac{\hat {f}_{coul} (\vec
{k},{\vec {k}}'')\hat {f}_{nucl} ({\vec {k}}'',{\vec {k}}''')\hat
{f}_{coul} ({\vec {k}}''',{\vec {k}}')}{(\varepsilon_{k} -
\varepsilon_{k'''} + i\eta )(\varepsilon_{k} -
\varepsilon_{k'''} + i\eta )}} d^3{\vec {k}}''d^3{\vec {k}}''' - .... \\
 \end{array}\eqno(2.10)
$$

\section{Polarization characteristics of the deuterons registered by
the detector during the beam's passage through a thin target}

\subsection{Scattering in a Thin Target}

 Due to a long-range character of Coulomb interaction, Coulomb scattering of deuterons by target nuclei
 occurs at small angles and the Coulomb amplitude for $\theta \ll 1$ is much larger than the nuclear scattering amplitude.
 In this case, for solving equation (1.18), one can apply the perturbation theory [11]
using as a zero approximation the solution of kinetic equation
(1.18), where the collision term is determined only by the Coulomb
interaction between the incident particle and the nuclei of
matter.
  The contribution to the evolution of the spin density matrix $\hat {\rho }(\vec
{k})$ coming from  nuclear scattering and Coulomb-nuclear interference  is considered as a correction.
It should be noted that taking account of nuclear and interference
factors as perturbations is valid only for such target thicknesses
$z$  for which multiple nuclear scattering can be neglected [10].
In this part, in order to illustrate the principal patterns of relationships  and simplify the form of obtained relations,
we shall further analyze the characteristics of a deuteron beam for the case of a very thin target, where the change in the
deuteron polarization state occurs only due to single scattering events in matter in addition to coherent scattering.

Let us consider the process of particle transmission through the target, whose thickness $z$ is much smaller than the mean
free path of the deuteron in matter, i.e., $z < 1 / N\sigma $, $\sigma $ is the total cross-section of the deuteron
scattering by a nucleus. As a consequence, in the first order perturbation theory, the solution of (1.19) can be represented in a form:

$$
\begin{array}{l}
\displaystyle\hat {\rho }(\vec {k},z) = \hat {\rho }(\vec {k},0) +
\frac{\pi i}{k}N\left[ {(\hat {f}(\vec {k},\vec {k}) + \hat
{f}^{\scriptscriptstyle+} (\vec {k},\vec {k})),\hat {\rho }(\vec
{k},0)} \right]z + \frac{\pi i}{k}N\left\{ {(\hat {f}(\vec
{k},\vec {k}) - \hat {f}^{\scriptscriptstyle+}(\vec {k},\vec
{k})),\hat {\rho }(\vec
{k},0)} \right\}z + \\[-10pt]
{}\\
\displaystyle
 + N\hat {f}(\vec {k},\vec {k}_0 )\hat {\rho }(0)\hat {f}^{\scriptscriptstyle+}(\vec {k}_0
,\vec {k})z,
\end{array}\eqno(3.1)
$$
where $\hat {\rho }(\vec {k},0)$ is the density matrix of the beam when it enters the target, i.e, when $z = 0$.
It describes the distribution over the momenta of the particles entering the target relative to the direction $\vec
{k}_0$.

In obtaining relation (3.1) it was also assumed that the initial angular distribution of the beam is much smaller
that the characteristic angular width of the differential scattering cross-section. In this case the amplitude
$\hat {f}(\vec {k},{\vec {k}}')$ in the term $\int {d\Omega _{\vec {k}'} \hat {f}(\vec {k},{\vec
{k}}')\hat {\rho }({\vec {k}}',0)\hat
{f}^{\scriptscriptstyle+}({\vec {k}}',\vec {k})} $ can be removed from the integrand at point  ${\vec {k}}'=\vec {k}_0$.
As a result, we have $\int
{d\Omega _{\vec {k}'} \hat {f}(\vec {k},{\vec {k}}')\hat {\rho
}({\vec {k}}',0)\hat {f}^{\scriptscriptstyle+}({\vec {k}}',\vec
{k})} \simeq \hat {f}(\vec {k},\vec {k}_0 )\hat {\rho }(0)\hat
{f}^{\scriptscriptstyle+}(\vec {k}_0 ,\vec {k})$, where $\hat {\rho
}(0) = \int{d\Omega _{\vec {k}'}\hat {\rho }({\vec {k}}',0)} $ is the spin part of the beam's density matrix
$\hat {\rho }(\vec {k},0)$.
The term $N\int {d\Omega _{\vec {k}'}} \hat {f}(\vec {k},{\vec
{k}}')\hat {\rho }({\vec {k}}',0)\hat
{f}^{\scriptscriptstyle+}({\vec {k}}',\vec {k})$ or $N\hat
{f}(\vec {k},\vec {k}_0 )\hat {\rho }(0)\hat
{f}^{\scriptscriptstyle+}(\vec {k}_0 ,\vec {k})$
is the contribution to the evolution of the density matrix, which describes single scattering of particles in the direction of $\vec {k}$.
It is known, in particular, that $\texttt{Sp}\hat {f}(\vec
{k},\vec {k}_0 )\hat {\rho }(0)\hat {f}^{\scriptscriptstyle+}(\vec
{k}_0 ,\vec {k})$ is the probability for a particle to undergo a
single elastic collision with a nucleus and get displaced by the
angle corresponding to the momentum direction $\vec {k}$
($\texttt{Sp}$ is taking the trace over the spin variables of the
deuteron).

Let us also consider the fact that scattering at high energies
chiefly occurs at small angles $\theta \ll 1$. The analysis shows
that in this case, the terms in the amplitude (2.1), which are
proportional to $B$ and  $C_1 $, lead to insignificant
depolarization of the detected beam [10] and will be dropped
hereinafter. As a result, the amplitude $\hat{f}$ in (3.1) has a
from:

$$
\hat {f}(\vec {k},{\vec {k}}') = f_0 (\theta ) + f_1 (\theta
)(\vec {S}\vec {n})^2,\eqno(3.2)
$$
where  $\vec {n} = \vec {k} / k$ is the unit vector in the direction of
$\vec {k}$, $f_0 = A -
2C_1 - 2C_2 $, $f_1 = 3C_2 $.

Using the solution (3.1) and the explicit form of the spin
structure of the amplitude $\hat {f}(\vec {k},{\vec {k}}')$ (3.2),
it is possible to find the dependence of the intensity and the
polarization characteristics of the beam on the direction of the
particle scattering and on the distance $z$ traveled by the
deuteron in matter. In a real experiment, the scattered particles
are registered  within a certain interval of finite momenta
because the collimator of the  detector has a finite angular
width.
We shall therefore consider further in this paper the characteristics of the beam transmitted
through the target in the interval of solid angles $\Delta \Omega $ with respect to the initial direction of the beam propagation.
In fact, due to the axial symmetry of the collimator, $\Delta \Omega $ is determined by the angular width of the detector collimator
$2\vartheta _{det} $ (for further calculations it is also assumed that $\vartheta _{det} $ is much larger than the initial angular
distribution of the beam).

\begin{figure}[!h]
\centering
\includegraphics[scale=0.7]{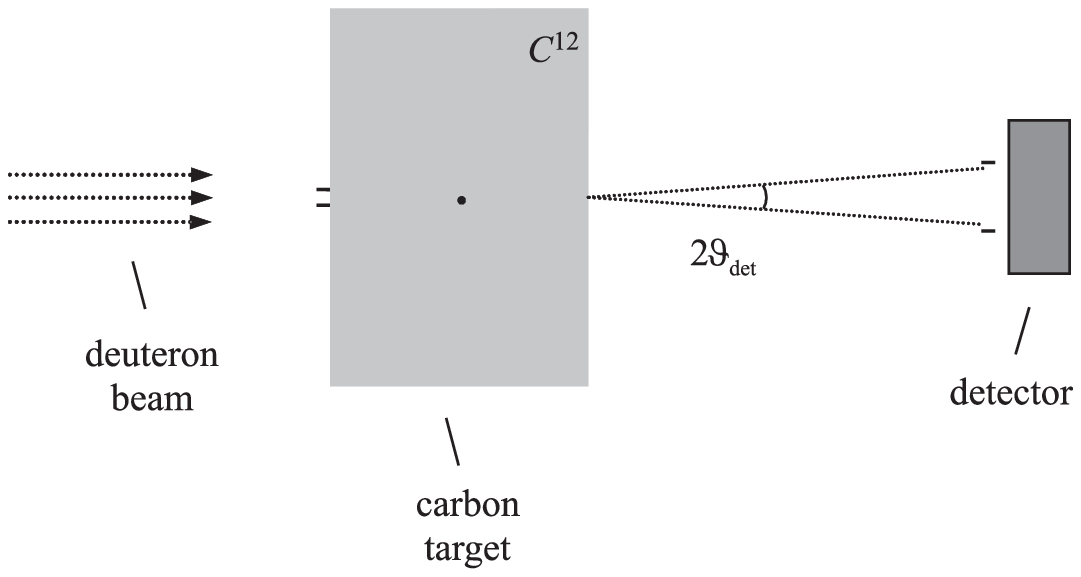}
\label{FigGr23}
\end{figure}

\begin{center}
{Fig. 1: Scheme of scattered beam detection  by the central
detector.}
\end{center}
\bigskip
In the case of deuterons (particles with spin 1), the polarization
state is characterized by the polarization vector $\vec {P}(\vec
{k}) =\texttt{Sp}\hat {\rho }(\vec {k})\hat {\vec {S}}$ and the
quadrupolarization tensor ${\rm {\bf Q}}$, whose components are
defined as $Q_{ik} (\vec {k}) = \texttt{Sp}\hat {\rho }(\vec
{k})\hat {Q}_{ik} $. The spin density matrix $\hat {\rho }(\vec
{k})$ can be written in the following general form:
$$\hat {\rho }(\vec {k}) = \frac{1}{3}I(\vec {k}){\hat {\mbox I}} +
\frac{1}{2}\vec {P}(\vec {k})\hat {\vec {S}} + \frac{1}{9}Q_{ik}
(\vec {k})\hat {Q}_{ik},\eqno(3.3)
$$

\noindent where $I(\vec {k}) = \texttt{Sp}\hat {\rho }(\vec {k})$.
Note that alongside with the quantities $\vec {P}$ and ${\rm {\bf
Q}}$, normalized spin characteristics of the beam are also used to
describe the beam polarization.

Denote the intensity, the polarization vector,  and the quadrupolarization tensor of the
 beam in the area occupied by the detector with angular width  $\Delta \Omega $ by ${\cal I}
\equiv \int_ {\scriptscriptstyle\Delta \Omega}\mathrm{d}\Omega
I(\vec{k},z)$, $\vec {\cal P} \equiv\int_
{\scriptscriptstyle\Delta
\Omega}\mathrm{d}\Omega\vec{P}(\vec{k},z)$, ${\cal Q} \equiv
\int_{\scriptscriptstyle\Delta \Omega}\mathrm{d}\Omega{\bf
Q}(\vec{k},z)$, respectively.
On the basis of the solution (3.1) of kinetic equation (1.18),
using the explicit form of the density matrix (3.3) and the
scattering amplitude (3.2), one can obtain the expression for the
integral characteristics of the deuteron beam:

$$
\begin{array}{l}
\displaystyle\mathcal{I}(z)=I_{0}+\displaystyle
N\left(-\sigma^{0}_{tot}+\int_{\scriptscriptstyle\Delta\Omega}
|f_{\scriptscriptstyle0}|^{2}\mathrm{d}\Omega\right)zI_{0}+\\[-10pt]
{}\\
\displaystyle
N\left(-(\sigma_{tot}^{\pm1}-\sigma_{tot}^{0})+2\int_{\scriptscriptstyle\Delta\Omega}
\mathrm{Re}(f_{\scriptscriptstyle0}f_{\scriptscriptstyle1}^{*})\mathrm{d}\Omega+
\int_{\scriptscriptstyle\Delta\Omega}|f_{\scriptscriptstyle1}|^{2}\mathrm{d}\Omega\right)\left[\frac{2}{3}I_{0}+\frac{1}{3}(\mathbf{Q}_{0}\vec{n})\vec{n}\right]z
,\\[-10pt]
{}\\
{}\\
\displaystyle\mathcal{\vec{P}}(z)=\vec{P}_{0}+
N\left(-\sigma^{0}_{tot}+\int_{\scriptscriptstyle\Delta\Omega}
|f_{\scriptscriptstyle0}|^{2}\mathrm{d}\Omega\right)z\vec{P}_{0}+
N\left(-\frac{1}{2}(\sigma_{tot}^{\pm1}-\sigma_{tot}^{0})+\int_{\scriptscriptstyle\Delta\Omega}
\mathrm{Re}(f_{\scriptscriptstyle0}f_{\scriptscriptstyle1}^{*})\mathrm{d}\Omega\right)\left[\vec{P}_{0}+\vec{n}(\vec{P}_{0}\vec{n})\right]z+\\[-10pt]
{}\\
\displaystyle
\left(\int_{\scriptscriptstyle\Delta\Omega}|f_{\scriptscriptstyle1}|^{2}d\Omega\right)\vec{n}(\vec{P}_{0}\vec{n})z-\frac{2}{3}\frac{2\pi
N}{k}\mathrm{Re}\left[f_{1}(0)-\frac{ik}{2\pi}\int_{\scriptscriptstyle\Delta\Omega}
f_{\scriptscriptstyle0}^{*}f_{1}\mathrm{d}\Omega\right][\vec{n}\times(\mathbf{Q}_{0}\vec{n})]z,\\[-10pt]
{}\\
{}\\
\displaystyle\mathcal{{Q}}(z)=\mathbf{Q}_{0}
+N\left(-\sigma_{tot}^{0}+\int_{\scriptscriptstyle\Delta\Omega}
|f_{\scriptscriptstyle0}|^{2}\mathrm{d}\Omega\right)z\mathbf{Q}_{0}+
N\left(-(\sigma_{tot}^{\pm1}-\sigma_{tot}^{0})+2\int_{\scriptscriptstyle\Delta\Omega}
\mathrm{Re}(f_{\scriptscriptstyle0}f_{\scriptscriptstyle1}^{*})\mathrm{d}\Omega\right)
\left[\frac{}{}\mathbf{Q}_{0}+\right.\\[-10pt]
{}\\
\displaystyle
\left.\frac{1}{3}(3\vec{n}\otimes\vec{n}-\mathbf{I})I_{0}-
\frac{1}{2}\left((\mathbf{Q}_{0}\vec{n})\otimes\vec{n}+\vec{n}\otimes(\mathbf{Q}_{0}\vec{n})\right)+
\frac{1}{3}\mathbf{I}(\mathbf{Q}_{0}\vec{n})\vec{n}\right]z+
\left(\int_{\scriptscriptstyle\Delta\Omega}|f_{\scriptscriptstyle1}|^{2}\mathrm{d}\Omega\right)\left[\frac{1}{3}(3\vec{n}\otimes\vec{n}-
\mathbf{I})I_{0}-\right.\\[-10pt]
{}\\
\displaystyle\left.
\frac{1}{2}\left((\mathbf{Q}_{0}\vec{n})\otimes\vec{n}+\vec{n}\otimes(\mathbf{Q}_{0}\vec{n})\right)+
+\frac{1}{2}\vec{n}^{\scriptscriptstyle\times}\mathbf{Q}_{0}\vec{n}^{\scriptscriptstyle\times}+\vec{n}\otimes\vec{n}\,\,(\mathbf{Q}_{0}\vec{n})\vec{n}
-\frac{1}{6}\mathbf{I}(\mathbf{Q}_{0}\vec{n})\vec{n}+\frac{1}{2}\mathbf{Q}_{0}\right]z-\\[-10pt]
{}\\
\displaystyle \frac{3}{2}\frac{2\pi
N}{k}\mathrm{Re}\left[f_{1}(0)-\frac{ik}{2\pi}\int_{\scriptscriptstyle\Delta\Omega}
f_{\scriptscriptstyle0}^{*}f_{1}d\Omega\right]\left([\vec{n}\times\vec{P}_{0}]\otimes\vec{n}+\vec{n}\otimes[\vec{n}\times\vec{P}_{0}]\right)z,
\end{array}\eqno(3.4)
$$
\noindent where $\vec {P}_0$ and ${\rm {\bf Q}}_0$ are the
polarization vector and the quadrupolarization tensor of the
deuteron beam at entering the target, respectively: $\vec {P}_0 =
\int {\mathrm{d}\Omega \vec {P}(\vec {k},0)}$, ${\rm {\bf Q}}_0 =
\int {\mathrm{d}\Omega {\rm {\bf Q}}(\vec {k},0)}$; $\vec {n} =
\vec {k}_0 / k_0 $ is the unit vector in the direction of the
deuteron momentum  $\vec {k}_0 $ (the quantization axis), $
\otimes $ is the dyadic product of vectors: $(\vec {n} \otimes
\vec {n})_{ij} = n_i n_j $, $\vec {n}^{\scriptscriptstyle\times} $
is the tensor dual to vector $\vec {n}$: $(\vec
{n}^{\scriptscriptstyle\times} )_{ij} = \varepsilon _{ijk} n_k $,
$({\rm {\bf Q}}_0 \vec {n})$ is the vector having the components
$({\rm {\bf Q}}_0 \vec {n})_l = Q_{0\,lk} n_k $.

Now let us analyze the obtained solutions. First of all, pay
attention to the fact that according to (3.4), the intensity and
the polarization characteristics of the beam depend on the
magnitude of the interval $\Delta \Omega $ of the solid angle. As
is seen, at $\Delta \Omega \to 0$, the contribution due to single
scattering disappears in expressions for ${\cal I}$, $\vec {\cal
P}$, and ${\cal Q}$.

Let us give a more detailed treatment of the change in the
polarization vector $\vec {\cal P}$ depending on the  target
thickness $z$. Let $\vec {P}_0 $ be an arbitrary, not equal to
$\pi / 2$, angle with the direction $\vec {n}$. Choose a
coordinate system so that the $z$-axis in it coincides with the
direction $\vec {n}$ of the deuteron incidence onto the target,
while the axes $x$  and $y$ are located in such a way that the
initial polarization vector $\vec {P}_0 $ lies in the $xz$ plane.

\begin{figure}[!h]
\centering
\includegraphics[scale=0.6]{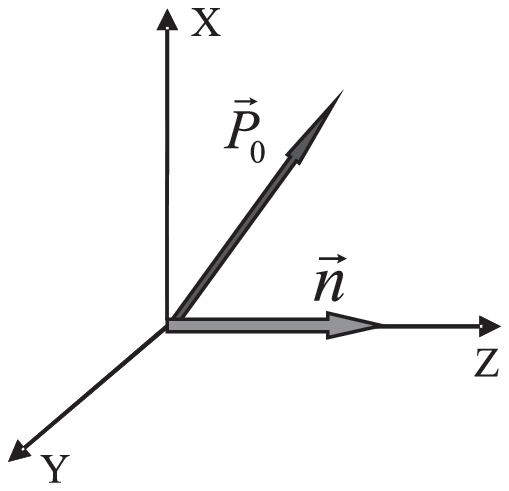}
\label{FigGr23}
\end{figure}
\begin{center}
{Fig. 2: Coordinate frame}
\end{center}
\bigskip

In this case $Q_{0\,yz}=0$, hence the polarization vector ${\rm
{\bf Q}}_0 \vec {n}$ also lies in the plane defined by vectors
$\vec {P}_0 $ and $\vec {n}$. Then from the expression for $\vec
{\cal P}$ follows that the first three terms in it describe the
change in the component of the polarization vector, which lies in
the plane  $(\vec {P}_0 ,\vec {n})$, i.e., these terms lead to the
change in the magnitude of the polarization vector
(depolarization).
The last term, which is defined by vector $[\vec {n}\times ({\rm {\bf Q}}_0 \vec
{n})]$, leads to the appearance in the initial polarization vector of a component perpendicular
to the plane $(\vec {P}_0 ,\vec {n})$, which grows with growing $z$, i.e., this term describes
the rotation of vector $\vec {\cal P}$ with respect to the direction of $\vec{n}$.  The rotation angle $\varphi _{eff}$ is defined as follows:

$$ \varphi_{eff}=\displaystyle\frac{2\pi
N}{k}\mathrm{Re}\left[f_{1}(0)-\frac{ik}{2\pi}\int_{\scriptscriptstyle\Delta\Omega}
f_{\scriptscriptstyle0}^{*}f_{1}\mathrm{d}\Omega\right]z.\eqno(3.5)
$$

As it was shown in [1, 2], the part of $\varphi _{eff} $, equal to
$\displaystyle\varphi_{0}\equiv\frac{2\pi
N}{k}\mathrm{Re}f_{1}(0)z$, is the rotation angle of the
polarization vector due to coherent scattering of the deuterons in
unpolarized matter.

When incoherent scattering by target nuclei is taken into account, then according to (3.5),
the rotation angle includes the additional contribution, which depends on  $\Delta
\Omega $.

To provide a more detailed treatment  of this dependence, it would be recalled that the
deuteron interacts with target nuclei via nuclear and Coulomb interactions (2.4).
The general spin structure of the corresponding amplitudes can be  represented as in (3.2).
This enables one to write the spinless and spin parts of the amplitude $\hat {f}$, which appear in (2.4) as follows:

$$f_0 (\theta ) = a(\theta ) + d(\theta ),\,\,\,\,\,f_1 (\theta ) =
d_1 (\theta ),\eqno(3.6)
$$
\noindent where $a(\theta )$ is the amplitude of the Coulomb
scattering of the deuteron by a nucleus with the charge $Z$ (the
amplitude in $f_1 (\theta )$ describing the spin part of the
Coulomb interaction is small in comparison with $d_1 (\theta )$
[19], and so it is assumed that $\hat {f}_{coul}(\theta ) =
a(\theta ){\hat {\mbox I}})$; $d(\theta )$ is the spin-independent
and  $d_1 (\theta )$ is the spin dependent parts of the modified
nuclear amplitude.
The explicit form of the dependence of $d$ on $\theta $ in the
limit of small scattering angles for the case of interaction of
structure particles was obtained, for example, in [20]. To
estimate the major parameters, let us consider the most simple
form of the dependence of $d$ and  $d_1 $ on $\theta $, namely:

$$d(\theta ) = d(0)e^{ - k^2R_d^2 \theta ^2 / 4},\,\,\,\,\,d_1
(\theta ) = d_1 (0)e^{ - k^2R_d^2 \theta ^2 / 4},\eqno(3.7)
$$
where $R_d $ is the deuteron radius.

\subsection{Contributions to the rotation angle due to incoherent
scattering}

Let us consider the terms in (3.4) which are integrated  over the
angular width $\Delta \Omega $ of the detector. According to
(3.6), the functions $f_0 (\theta )$ and $f_1 (\theta )$ included
in the obtained solutions contain the contributions from the
Coulomb and nuclear amplitudes, which, in turn, have different
angular dependence. As a result, in analyzing (3.4)  for the case
of high energies, we shall select such limits of $\vartheta _{det}
$ that are bounded by the diffraction angles $\theta _c $ and
$\theta _n $ of scattering of fast deuterons, for Coulomb and
nuclear scattering, respectively.
For a screened Coulomb potential, $\theta _c \sim 1 / kR_c$, where $R_c $ is the shielding radius
(in the case of scattering by a carbon target, $R_c = 3 \cdot 10^{ - 9}$ cm). When deuterons are scattered
by nuclei whose radius is smaller than that of the deuterons, $\theta _n \sim 1 / kR_d $.

 In view of the representation of the  amplitude $\hat {f}(\theta )$ in (3.6), we obtain the following general expression
 for the rotation angle of the polarization vector (3.5):

$$ \varphi_{eff}=\displaystyle\frac{2\pi
N}{k}\mathrm{Re}\left[d_{1}(0)-\frac{ik}{2\pi}\int_{\scriptscriptstyle\Delta\Omega}
a^{*}(\theta)d_{1}(\theta)\mathrm{d}\Omega-\frac{ik}{2\pi}\int_{\scriptscriptstyle\Delta\Omega}
d^{*}(\theta)d_{1}(\theta)\mathrm{d}\Omega\right]z.\eqno(3.8)
$$
where the dependence of the nuclear amplitude on the scattering angle for high energies is determined by (3.7).
For further analysis of the magnitude of the effect under consideration, it is convenient to represent $\varphi
_{eff} $  in the form
$$\varphi _{eff} (\vartheta _{det} ) = \varphi _0
+ \varphi _{nc} (\vartheta _{det} ) + \varphi _{nn} (\vartheta
_{det} ),\eqno(3.9)
$$

\noindent where $\displaystyle\varphi _0 = \frac{2\pi
N}{k}\mbox{Re}d_1 (0)z$ is the contribution of the coherent scattering to the rotation angle of the
polarization vector; $\displaystyle\varphi_{nc}(\vartheta _{det})=
N\int_{\scriptscriptstyle\Delta
\Omega}\mathrm{Im}\left[a^{*}(\theta)d_{1}(\theta)\right]\mathrm{d}\Omega
z$ is the part of the rotation angle, which describes the effect of the Coulomb-nuclear interference;
$\displaystyle\varphi_{nn}(\vartheta _{det})=
N\int_{\scriptscriptstyle\Delta
\Omega}\mathrm{Im}\left[d^{*}(\theta)d_{1}(\theta)\right]\mathrm{d}\Omega
z$ is the correction to the deuteron spin rotation from  nuclear scattering.

Numerical values of these contributions obtained, for example, for the energy of 500 MeV ($k = 0.74 \cdot 10^{14}$ cm$^{ - 1}$).
Then in the case of scattering by a carbon target, we have $\theta _c \sim 10^{ -5}$rad, $\theta _n \sim 10^{ - 2}$rad.

For the imaginary and real parts of the nuclear amplitude of
deuteron scattering by a carbon nucleus at zero angle, take the
values calculated in the eikonal approximation in the energy
region  $E = 0.5\div 1$ GeV: $f_{M = 0}^{nucl} (0) = - 5.97 +
74.81i\,\,$fm, $f_{M = \pm 1}^{nucl} (0) = - 5.84 + 73.9i\,\,$fm
[21]. From this $\mbox{Im}d = 0.75 \cdot 10^{ - 11}$cm,
$\mbox{Re}d = - 0.6 \cdot 10^{ - 12}$cm, $\mbox{Im}d_1 = - 0.91
\cdot 10^{ - 13}$cm, $\mbox{Re}d_1 = 0.13 \cdot 10^{ - 13}$cm. As
a result, we have  $\varphi _0 \simeq 1.2 \cdot 10^{ - 4}z$.

Let us consider the ratio ${\varphi _{nc} } \mathord{\left/
{\vphantom {{\varphi _{nc} } {\varphi _0 }}} \right.
\kern-\nulldelimiterspace} {\varphi _0 }$.
Perform integration over the angular width of the detector, using
the expression for the Coulomb amplitude in the first Born
approximation. This is possible because for the deuteron energies
considered here, in scattering by  a screened Coulomb potential,
the ratio of the real part of the amplitude in the second-order
perturbation theory  approximation $a^{(2)}$ to that in the Born
approximation  is
$\displaystyle\left|\mathrm{Re}a^{(2)}/\mathrm{Re}a^{(1)}\right|\sim\frac{mZe^{2}}{\hbar^{2}k}\frac{1}{k
R_{c}}\ll 1$ for the whole range of scattering angles.

For the range of angles $\theta \ll \theta _c $, the ratio is
$\displaystyle\left|\mathrm{Im} a^{(2)}/\mathrm{Re} a^{(1)}
\right| \sim \frac{mZe^2}{\hbar ^2k} = 0.057$; this estimate also
remains unchanged with increasing $k\theta R_c$ [22].
For $k\theta R_c \gg 1$ the ratio $\displaystyle\mathrm{Im}
a^{(2)} / \mathrm{Re} a^{(1)} \simeq - 2\frac{mZe^2}{\hbar ^2k}\ln
(k\theta R_c )$. That is why for, e.g., $\theta \sim \theta _n $,
this ratio is of the order of unity. Taking into account the
magnitude of $\left| {\mbox{Im}d_1 / \mbox{Re}d_1 } \right| \simeq
7$, as well as a fast decrease in the integrand due to the nuclear
amplitude, one can demonstrate that for $\theta \gtrsim\theta _n
$, the contribution to ${\varphi _{nc} } \mathord{\left/
{\vphantom {{\varphi _{nc} } {\varphi _0 }}} \right.
\kern-\nulldelimiterspace} {\varphi _0 }$ of the term containing
$\mathrm{Im} a(\theta )$ will be an order of magnitude smaller
than the magnitude of ${\varphi _{nc} } \mathord{\left/ {\vphantom
{{\varphi _{nc} } {\varphi _0 }}} \right.
\kern-\nulldelimiterspace} {\varphi _0 }$, calculated using the
Coulomb amplitude in Born approximation. It should be emphasized
here that this term should not be ignored in those rare cases,
when $\left| {\mbox{Im}d_1 / \mbox{Re}d_1}\right| \sim 1$, as well
as in the cases of calculating the cross-section of
Coulomb-nuclear interaction and spin dichroism (for calculations
in the case of high energies $\displaystyle\frac{mZe^2}{\hbar ^2k}
\ll 1$, it is sufficient to use the imaginary part of the Coulomb
amplitude calculated in the second order perturbation theory).

Thus, in  the whole range of
scattering angles one can obtain for $\varphi _{nc} / \varphi _0$:

$$
\frac{\varphi _{nc}}{\varphi _0 }= - \frac{\mathrm{Im} d_1
}{\mathrm{Re} d_1 }\,\frac{mZe^2}{\hbar ^2k}\left\{
{\mbox{Ei}\left( { - \frac{R_d^2 }{4R_c^2 } - \frac{k^2R_d^2
\vartheta _{det}^2 }{4}} \right) - \mbox{Ei}\left( { - \frac{R_d^2
}{4R_c^2 }} \right)} \right\},\eqno(3.10)
$$
\noindent where $\displaystyle\mbox{Ei}( - z) =
-\int^{\infty}_{z}\frac{\mathrm{d}x}{x}e^{-x}$ is the integral
exponential function. The relative contribution to $\varphi _{nn}
/ \varphi _0 $ of nuclear scattering of the deuterons by the
target nuclei is obtained, using the approximate expression (3.7):
$$
\displaystyle\frac{\varphi _{nn} }{\varphi _0 } = \frac{1}{kR_d^2
}\left\{ {\mathrm{Re} d\frac{\mathrm{Im} d_1 }{\mathrm{Re} d_1 } -
\mathrm{Im} d} \right\}\left( {1 - e^{ - R_d^2 k^2\vartheta
_{det}^2 / 2}} \right).\eqno(3.11)
$$

The analysis shows that for the range of values   $\vartheta
_{det} \ll 1 / kR_c $ the relationships  (3.10) and (3.11) are
much smaller than unity, i.e., the rotation angle of the
polarization vector in this case is determined by coherent
scattering of particles in matter (is determined by  refraction of particles in matter).

Note also that for stated $\vartheta _{det} $, the beam
depolarization and spin dichroism will only be determined by the
total cross-section $\sigma _{tot}^0 $ and $\sigma _{tot}^{\pm 1}
$.

As a result, for $\vartheta _{det} \ll 10^{ - 5}$, the system of
solutions (3.4) is
reduced to the solutions describing the evolution of the spin state of the deuteron beam only due to particle refraction in the target.
These solutions were obtained in [1, 2].%\cite{1,2}.

With increasing $\vartheta _{det} $ (even for $\vartheta _{det}
\sim 1 / kR_c )$, singly scattered particles start affecting the
polarization characteristics of the beam. For instance, for
$\vartheta _{det} \sim 10^{ - 3}$ rad, the contribution of the
Coulomb-nuclear interference to the rotation angle $\varphi _{nc}
$ of the polarization vector of the deuteron beam is of the order
of $\varphi _0 $, moreover, with growing $\vartheta _{det} $
($\vartheta _{det} \ll \theta _n )$, according to (3.10),
$\varphi _{nc} $ grows logarithmically: ${\varphi _{nc} }
\mathord{\left/ {\vphantom {{\varphi _{nc} } {\varphi _0 }}}
\right. \kern-\nulldelimiterspace} {\varphi _0 } \sim \ln (kR_c
\vartheta _{det} )$.

\subsection{Rotation angle including the Coulomb contribution to
the spin part of the nuclear amplitude $d_{1}(0)$}.

It should be noted that if in (3.8), integration over
$\mathrm{d}\Omega $ is made over the entire solid angle (a $4\pi$
experimental geometry),
then according to the analysis given in [11, 12], %\cite{9,10},
the considered interference term and the contribution to the
amplitude $d_1(0)$ due to distortion of the incident waves by
Coulomb interaction compensate one another, i.e., the sum of the
contributions $\varphi _0 + \varphi _{nc} (\vartheta _{det} = \pi
)$ is in this case determined by the spin-dependent part of a pure
nuclear amplitude of scattering at zero angle $\displaystyle
{\varphi }'_0 \equiv \varphi _0 + \varphi _{nc} (\pi ) =
\frac{2\pi N}{k}\mathrm{Re} d_1 '$. This will be demonstrated
below.

The modified nuclear amplitude  $\hat {f}_{nucl,coul} (\vec
{k},{\vec {k}}')$ is associated with "pure"\, Coulomb and "pure"\,
nuclear amplitudes by formula (2.10). Taking into account that the
solutions of (3.4) and  $\varphi _{eff} $ include only the
spinless part of the scattering amplitude $a(\theta )$, according
to (2.9), the spin-dependent part of $d_1 $ can be written in the
form:
$$
d_1 (\vec {k},{\vec {k}}') \simeq d_1 ^\prime (\vec {k},{\vec
{k}}') - \frac{1}{(2\pi )^2m}\int {\frac{d_1 ^\prime (\vec
{k},{\vec {k}}'')a({\vec {k}}'',{\vec {k}}')}{\varepsilon_{k} -
\varepsilon_{k''} + i\eta }} d^3{\vec {k}}'' - \frac{1}{(2\pi
)^2m}\int {\frac{a(\vec {k},{\vec {k}}'')d_1 ^\prime ({\vec
{k}}'',{\vec {k}}')}{\varepsilon_{k} - \varepsilon_{k''} + i\eta
}} d^3{\vec {k}}''.\eqno(3.12)
$$
The amplitude $d_1'$ is a "pure"\, nuclear spin-dependent
amplitude of scattering at angle $\theta $.

Substituting the explicit form of the amplitude $d_1 (\vec
{k},{\vec {k}}')$ from  (3.12) into the expression for $\varphi
_{eff} $ in (3.8), and again taking into account only the first
degree of the product of $d_1'$ and $a$, one  obtains

$$
\varphi _{eff}=\frac{2\pi
N}{k}\mathrm{Re}\left[d_{1}'(0)+\frac{ik}{2\pi}\int
a(\theta)d_{1}'(\theta)\mathrm{d}\Omega-\frac{ik}{2\pi}\int_{\scriptscriptstyle\Delta\Omega}
a^{*}(\theta)d_{1}'(\theta)\mathrm{d}\Omega-\frac{ik}{2\pi}\int_{\scriptscriptstyle\Delta\Omega}
d^{*}(\theta)d_{1}'(\theta)\mathrm{d}\Omega\right]z.\eqno(3.13)
$$
Integration in (3.12) was carried out subject to the following
representation of the stationary Green's function:
$$\frac{1}{\varepsilon_{k} - \varepsilon_{k'} + i\eta } = P\frac{1}{\varepsilon_{k} -
\varepsilon_{k'}} - i\pi \delta (\varepsilon_{k} -
\varepsilon_{k'}),\eqno(3.14)
$$
\noindent where $P$ is the principal value integration.

As is seen from (3.13), when the detector registers particles in
the $4\pi$ solid angle geometry ($\Delta\Omega=4\pi$), the
Coulomb-nuclear contributions are mutually canceled. The
compensation, however, only occurs when we can confine ourselves
to the consideration of the Coulomb scattering amplitude in the
first Born approximation.

Allowing for the Coulomb contribution to $d_1 (0)$, equation (3.9)
can be written as:

$$\varphi _{eff} = \varphi _0 ' + \varphi _{nc}^{tot} +
\varphi _{nn},\eqno(3.15)
$$
\noindent where $\varphi _{nc}^{tot}$ is the total contribution of
the Coulomb-nuclear interference to the rotation angle of the
polarization vector. Its magnitude is determined by the sum of the
second and third terms in (3.13). Within the limits of small
scattering angles, the explicit form of $\varphi _{nc}^{tot}$
reads as follows:
\[
\displaystyle\varphi _{nc}^{tot}= 2\pi Nz\int_{\vartheta _{det}
}^\infty \mathrm{Im}\left[ {a(\theta){d}'_1 (\theta )}
\right]\theta \mathrm{d}\theta=- 2\pi N \mathrm{Im}{d}'_1
\frac{mZe^2}{\hbar ^2 k^2}\mbox{Ei}\left( { - \frac{R_d^2 }{4R_c^2
} - \frac{k^2R_d^2 \vartheta _{det}^2 }{4}} \right)z.
\]

In this case the relative contribution of the Coulomb-nuclear interference to the rotation angle is determined by formula
$$\frac{\varphi
_{nc}^{tot} }{{\varphi }'_0 } = - \frac{\mathrm{Im}{d}'_1
}{\mathrm{Re}{d}'_1 }\frac{mZe^2}{\hbar ^2k^2}\mbox{Ei}\left( { -
\frac{R_d^2 }{4R_c^2 } - \frac{k^2R_d^2 \vartheta _{det}^2 }{4}}
\right)\eqno(3.16)
$$
For high energies, when nuclear scattering has a pronounced
diffraction character, one can estimate the magnitude of the lower
limit of the angle $\vartheta _{det}^{\scriptscriptstyle comp} $,
at which the interference contributions can be considered to
compensate one another with the selected accuracy. We have that
for the considered energy and the form of nuclear amplitude, the
total contribution of the Coulomb-nuclear interference can be
neglected even at $\vartheta _{det} \sim \theta _n$ (the
diffraction angle of nuclear scattering) ($\vert \varphi
_{nc}^{tot} \vert $ is an order of magnitude smaller than $\varphi
_0 ^\prime )$.

For example, if $\vartheta _{det} = 0.8 \cdot 10^{
- 2}$ rad,  $\varphi _{nc}^{tot} / \varphi _0 ^\prime = - 0.2$.
From this it can be assumed that $\vartheta
_{det}^{\scriptscriptstyle comp} \sim \theta _n $.

If the
collimator of the detector registers particles moving within a
certain solid angle $\Delta \Omega $ corresponding to $\vartheta
_{det} \ll \vartheta _{det}^{\scriptscriptstyle comp} $, the
stated Coulomb-nuclear terms will not be compensated. For example,
when $\vartheta _{det} \simeq 0.4 \cdot 10^{ - 3}$ rad, the part
of the rotation angle $\varphi _{nc}^{tot} $ is comparable in
magnitude with $\varphi _0 ^\prime $.

It should be emphasized that the minimum value of $\vartheta
_{det} $, when the Coulomb-nuclear contributions are compensated
($\vartheta _{det}^{\scriptscriptstyle comp}$), depends on the
deuteron energy ($ \sim 1 / \sqrt E )$ as well as on the type of
the dependence of the spin part of the nuclear amplitude on the
scattering angles.

Consider now the nuclear contribution to the rotation of the
deuteron spin. The estimate of  relation (3.11) for $\vartheta
_{det} \sim \theta _n $ gives $\varphi _{nn} / \varphi _0 \simeq -
10^{ - 2}$. With growing $\vartheta _{det} $, namely for
$\vartheta _{det} \gg 1 / kR_d$, the ratio $\vert \varphi _{nn} /
\varphi _0 \vert $ achieves its maximum value: $\vert \varphi
_{nn} / \varphi _0 \vert \simeq 0.3$, while the magnitude of the
ratio $\vert \varphi _{nn} / \varphi _0 ^\prime \vert $ is of the
order of $10^{ -2}$.

Let us mention the following possibility of experimental measuring
of the spin-dependent part of the zero-angle nuclear amplitude.
Measuring the rotation angle of the deuteron polarization vector
for two arbitrary values of the angles $\vartheta _1 $ and
$\vartheta _2$ of the detector and considering the difference
$\varphi _{eff} (\vartheta _1 ) - \varphi _{eff} (\vartheta _2 )$,
one obtains that $\varphi _{eff} (\vartheta _1 ) - \varphi _{eff}
(\vartheta _2 ) \approx \varphi _{nc} (\vartheta _1 ) - \varphi
_{nc} (\vartheta _2 )$.
Using equation (3.10), one can thus find the value of $\mathrm{Im}
d_1 $ and, hence the value of $\varphi _{nc} $ for any $\vartheta
_{det} $.

Knowing them and taking into account that $\left| {\varphi _{nn} } \right|
\ll \left| {\varphi _{nc} } \right|$ for any  $\vartheta _{det}$, it is possible, using (3.9),
to estimate the magnitude of the rotation angle $\varphi _0 $ due coherent scattering and, hence, the magnitude  of $\mathrm{Re} d_1 $.
This part of $d_1(0)$ can also be obtained  directly measuring the
rotation angle of the polarization vector for $\vartheta _{det}
\ll 1 / kR_c $.
Basing on the assumption that the Coulomb-nuclear contributions in
$\varphi _{eff} $ are compensated and choosing the angle of the
detector so that $\vartheta _{det} \gg \vartheta
_{det}^{\scriptscriptstyle comp} $, one can obtain the magnitude
of $\mathrm{Re}d_1 ^\prime $.

The domain of applicability of the solutions of (3.4) corresponds to such target thicknesses $z$ for
which multiple scattering in matter can be neglected, i.e.,  $z$ is much smaller
than the mean free path of the deuterons due to Coulomb scattering:
$z \le 1 / N\sigma _{coul} $.
For energies $\mbox{0.5}\div $1 GeV,  thickness $z$ is of the
order of $10^{ - 5}$ cm.

In the next section we shall demonstrate the validity of the
assertion that in the cases when $z$ is much larger than the mean
free path $1 / N\sigma _{coul} $ (i.e., in the targets where
multiple Coulomb scattering takes place), the contribution of the
Coulomb-nuclear interference to the rotation angle of the
polarization vector also depends on the angle of the detector
collimator.

$\\[-10pt]
{}\\$

\section{Polarization characteristics of deuterons under the
conditions of multiple Coulomb scattering in matter}

With increasing target thickness, at least the condition of the
smallness of the mean free path due to Coulomb interaction
$N\sigma _{coul} z \ll 1$ is  violated.  In this case the solution
of the kinetic equation, which  describes the Coulomb scattering
of a deuteron by the nuclei of matter should be found. It
will describe the beam distribution due to single and multiple
Coulomb collisions of particles in the target.

As has been stated
above, if in this case  $z \ll 1 / N\sigma _{nucl} $, which is
realized in the majority of practical cases, one can find the
nuclear and Coulomb nuclear contributions in the scope of the
perturbation theory.

\subsection{Solution of the kinetic equation}

Write equation (1.18) as follows:
$$
\begin{array}{l}
\displaystyle \frac{d\hat {\rho }(\vec {k})}{dz} = - N\sigma
_{tot}^0 \hat {\rho }(\vec {k}) - \frac{N}{2}\left( {\sigma
_{tot}^{\pm 1} - \sigma _{tot}^0 } \right)\left\{
{(\vec{S}\vec{n})^{2},\hat {\rho }(\vec {k})} \right\} +
\frac{2\pi i}{k}N \mathrm{Re} f_1 (0)\left[
{(\vec{S}\vec{n})^{2},\hat {\rho }(\vec {k})} \right] + \\[-10pt]
{}\\
 \displaystyle + N\int {d\Omega _{\vec {k}'} \hat {f}(\vec {k},{\vec {k}}')\hat {\rho
}({\vec {k}}')\hat {f}^{\scriptscriptstyle +}(\vec {k}',{\vec
{k}})}
 \end{array}\eqno(4.1)
$$
\noindent where $\sigma _{tot}^{\scriptscriptstyle 0,\pm 1} =
Sp\int {\mathrm{d}\Omega _{\vec {k}'} \hat {f}(\vec {k},{\vec
{k}}')\hat {\rho }_0^{\scriptscriptstyle 0,\pm 1} } \hat
{f}^{\scriptscriptstyle +} (\vec {k},{\vec
{k}}')+\sigma_{r}^{\scriptscriptstyle 0,\pm 1}$, $\hat {\rho
}_0^{\scriptscriptstyle 0,\pm 1}$ is the deuteron density matrix
describing the initial state of the beam for $M = 0$ and $M = \pm
1$, correspondingly; $\vec{n}$ is the unit vector in the direction
of  $\vec{k}$; $\sigma_{r}$ denotes the part of the total
cross-section, which is responsible for all possible inelastic
interactions between the deuteron  and the nuclei of matter,
including nuclear reactions.
Substituting (2.4) into the expressions for the total
cross-sections, the elastic part of $\sigma
_{tot}^{\scriptscriptstyle 0,\pm 1} $ can be represented as a sum
of the terms describing the cross-section of the  Coulomb
scattering, the pure nuclear cross-section and the cross-section
of the interference between the Coulomb and nuclear interactions:
$\sigma _{tot}^{\scriptscriptstyle 0,\pm 1} = \sigma
_{coul}^{\scriptscriptstyle 0,\pm 1} + \sigma
_{nucl,coul}^{\scriptscriptstyle 0,\pm 1} + \sigma
_{nucl}^{\scriptscriptstyle 0,\pm 1}+
\sigma_{r}^{\scriptscriptstyle 0,\pm 1}$.

Use the following
notations: $\sigma _{NC}^{\scriptscriptstyle 0,\pm 1} = \sigma
_{nucl,coul}^{\scriptscriptstyle 0,\pm 1} + \sigma _{nucl}^{0,\pm
1}+\sigma_{r}^{\scriptscriptstyle 0,\pm 1}$. Thus, represent the
cross-sections $\sigma _{tot}^{\scriptscriptstyle 0,\pm 1} $ in
the form:

$$\sigma _{tot}^{\scriptscriptstyle 0,\pm 1} = \sigma _{coul}^{\scriptscriptstyle 0,\pm 1} + \sigma
_{NC}^{\scriptscriptstyle 0,\pm 1},\eqno(4.2)
$$
\noindent where $\sigma _{coul}^{\scriptscriptstyle 0,\pm 1} =
Sp\int {d\Omega _{\vec {k}'} \hat {f}_{coul} (\vec {k},{\vec
{k}}')\hat {\rho }_0^{\scriptscriptstyle 0,\pm 1} } \hat
{f}_{coul}^{\scriptscriptstyle +} ({\vec {k}}',\vec {k}),$ $\sigma
_{NC}^{\scriptscriptstyle 0,\pm 1} = Sp\int {d\Omega _{\vec {k}'}
\left( {\hat {f}_{coul} (\vec {k},{\vec {k}'})\hat {\rho
}_0^{\scriptscriptstyle 0,\pm 1} \hat
{f}_{nucl,coul}^{\scriptscriptstyle +}({\vec {k}}',\vec
{k})}\right.}+$

$
 \left.{+ \hat {f}_{nucl,coul} (\vec {k},{\vec {k}}')\hat
{\rho }_0^{\scriptscriptstyle 0,\pm 1} \hat
{f}_{coul}^{\scriptscriptstyle +} ({\vec {k}}',\vec {k}) +\hat
{f}_{nucl,coul} (\vec {k},{\vec {k}}')\hat {\rho
}_0^{\scriptscriptstyle 0,\pm
1} \hat {f}_{nucl,coul}^{\scriptscriptstyle +} ({\vec {k}}',\vec {k})} \right)+\sigma_{r}^{\scriptscriptstyle 0,\pm 1}. \\
$

In view of the notations introduced above and the representation
of the scattering amplitude (2.4), rewrite kinetic equation (4.1)
as:

$$
\begin{array}{l}
 \displaystyle\frac{d\hat {\rho }(\vec {k})}{dz} = - N\sigma _{coul}^{\scriptscriptstyle 0} \hat {\rho }(\vec
{k}) - \frac{N}{2}\left( {\sigma _{coul}^{\scriptscriptstyle \pm
1} - \sigma _{coul}^{\scriptscriptstyle 0} } \right)\left\{
{(\vec{S}\vec{n})^{2},\hat {\rho }(\vec {k})} \right\} +
\frac{2\pi i}{k}N \mathrm{Re} f_1 ^{coul}(0)\left[ {(\vec{S}\vec{n})^{2},\hat {\rho }(\vec {k})} \right] + \\[-10pt]
{}\\
 \displaystyle
 + N\int {d\Omega _{\vec {k}'} \hat {f}_{coul} (\vec {k},{\vec {k}}')\hat
{\rho }({\vec {k}}')\hat {f}^{\scriptscriptstyle +}_{coul} ({\vec {k}}',\vec {k})} - \\[-10pt]
{}\\
 \displaystyle
 - N\sigma _{NC}^{\scriptscriptstyle 0} \hat {\rho }(\vec {k}) - \frac{N}{2}\left( {\sigma
_{NC}^{\scriptscriptstyle\pm 1} - \sigma
_{NC}^{\scriptscriptstyle0} } \right)\left\{
{(\vec{S}\vec{n})^{2},\hat {\rho }(\vec {k})} \right\} +
\frac{2\pi i}{k}N \mathrm{Re} f_1 ^{nucl}(0)\left[
{(\vec{S}\vec{n})^{2},\hat {\rho }(\vec {k})}
\right] + \\[-10pt]
{}\\
 \displaystyle
 + N\int {d\Omega _{\vec {k}'} \left( {\hat {f}_{coul} (\vec {k},{\vec
{k}}')\hat {\rho }({\vec {k}}')\hat {f}^{\scriptscriptstyle
+}_{nucl,coul} ({\vec {k}}',\vec {k}) + \hat {f}_{nucl,coul} (\vec
{k},{\vec {k}}')\hat {\rho }({\vec
{k}}')\hat {f}^{\scriptscriptstyle +}_{coul} ({\vec {k}}',\vec {k})} \right)} + \\[-10pt]
{}\\
 \displaystyle
 + N\int {d\Omega _{\vec {k}'} \hat {f}_{nucl,coul} (\vec {k},{\vec
{k}}')\hat {\rho }({\vec {k}}')\hat {f}^{\scriptscriptstyle
+}_{nucl,coul} ({\vec {k}}',\vec {k}),}
 \end{array}\eqno(4.3)
$$

As stated above, the solution of equation (4.3) can be presented as follows:

$$\hat {\rho }(\vec {k},z) = \hat {\rho }^{(0)}(\vec
{k},z) + \hat {\rho }^{(1)}(\vec {k},z) + ...,\eqno(4.4)
$$
\noindent where $\hat {\rho }^{(0)}(\vec {k},z)$ is the zero
approximation, which is the solution of kinetic equation (4.3) and
describes only the Coulomb interaction between the deuteron beam
and the nuclei of matter; $\hat {\rho }^{(1)}(\vec{k},z)$ is the
first order perturbation theory correction allowing for the
contribution of nuclear scattering to the evolution of the
polarization characteristics of the beam.

The equation for $\hat {\rho }^{(0)}(z)$ has a form:

$$
\begin{array}{l}
 \displaystyle\frac{d\hat {\rho }^{(0)}(\vec {k})}{dz} = - N\sigma _{coul}^{\scriptscriptstyle 0} \hat {\rho
}^{(0)}(\vec {k}) - \frac{N}{2}\left( {\sigma
_{coul}^{\scriptscriptstyle\pm 1} - \sigma
_{coul}^{\scriptscriptstyle 0} } \right)\left\{
{(\vec{S}\vec{n})^{2},\hat {\rho }^{(0)}(\vec {k})} \right\} +
\frac{2\pi i}{k}N \mathrm{Re} f_1 ^{coul}(0)\left[ {(\vec{S}\vec{n})^{2},\hat {\rho }^{(0)}(\vec {k})} \right] + \\[-10pt]
{}\\
 \displaystyle
 + N\int {d\Omega _{\vec {k}'} \hat {f}_{coul} (\vec {k},{\vec {k}}')\hat
{\rho }^{(0)}({\vec {k}}')\hat {f}^{\scriptscriptstyle +}_{coul} (\vec {k},{\vec {k}}')} , \\
 \end{array}\eqno(4.5)
$$
\noindent where the spin structure $\hat {f}_{coul} (\vec
{k},{\vec {k}}')$ in the general case has a form (2.1).

It can be shown [10] that the terms in the Coulomb amplitude
(2.1), which are proportional to $B$, $C_1 $, and to the part of
$C_2 $ depending on $\theta $, lead to depolarization of the
registered beam. The magnitude of such depolarization is
determined by $b_g^2 \overline {\theta ^2} z$ ($\displaystyle b_g
= \frac{g - 2}{2}\frac{\gamma ^2 - 1}{\gamma } + \frac{\gamma -
1}{\gamma }$, $g$ is the gyromagnetic ratio, $\displaystyle
\overline {\theta ^2} = N\int {\theta ^2\frac{d\sigma _{coul}
}{d\Omega }} d\Omega$ is the average squared angle of Coulomb
scattering).
 With due account of the small values of the scattering angles due to Coulomb scattering at high energies,
 this quantity will be insignificant as compared with the contribution from the scalar part of the amplitude (2.1).

 For this reason, in the case of Coulomb interaction we can confine ourselves to considering the scattering amplitude of the form  (3.2).

 Moreover, the term in (3.2), which is equal to $f_1 ^{coul}(0)(\vec {S}\vec {n})^2$ and describes a
 coherent rotation of the polarization vector of the deuteron beam due to the elastic Coulomb
 scattering at zero angle, is also small and  thus can be neglected [19].

As a result, equation (4.5) is rewritten as follows:

$$\frac{d\hat {\rho }^{(0)}(\vec {k},z)}{dz} = - N\sigma _{coul}^{\scriptscriptstyle0}
\hat {\rho }^{(0)}(\vec {k},z) + N\int {d\Omega _{\vec {k}'}
\left| {a(\vec {k},{\vec {k}}')} \right|^2\hat {\rho }^{(0)}({\vec
{k}}',z)},\eqno(4.6)
$$
\noindent where $a(\vec {k},{\vec {k}}')$ denotes the spinless part of the amplitude $\hat {f}_{coul} (\vec {k},{\vec
{k}}')$.

Expression (4.6) is an integro-differential equation. Its solution
in the limit of small scattering angles can be obtained by
expanding the function $\hat {\rho }^{(0)}({\vec {k}}')$ over a
relatively small parameter $\vec {q}$ (transferred momentum) [23]:

$$\hat {\rho }^{(0)}({\vec {k}}',z) \approx \hat {\rho }^{(0)}(\vec
{k},z) + \frac{\partial \hat {\rho }^{(0)}}{\partial k_x }q_x +
\frac{\partial \hat {\rho }^{(0)}}{\partial k_y }q_y +
\frac{1}{2}\frac{\partial ^2\hat {\rho }^{(0)}}{\partial k_x
^2}q_x ^2 + \frac{\partial ^2\hat {\rho }^{(0)}}{\partial k_x
\partial k_y }q_x q_y + \frac{1}{2}\frac{\partial ^2\hat {\rho
}^{(0)}}{\partial k_y ^2}q_y ^2 + ...,\eqno(4.7)
$$

\noindent where $q_x = q\cos \varphi $, $q_y = q\sin \varphi $, $q
\simeq k\theta $, where $\theta$ is the scattering angle (the angle between vector $\vec {k}$ and the $z$-axis).

Substituting the expansion (4.7) into (4.6) and integrating over
the azimuth angle $\varphi $, one can obtain

$$ \frac{d\hat {\rho }^{(0)}(\vec {k},z)}{dz} = \frac{\overline
{\,\theta ^2} }{4}\Delta \hat {\rho }^{(0)}(\vec {k},z) +
\frac{\overline {\,\theta ^4} }{64}\Delta \Delta \hat {\rho
}^{(0)}(\vec {k},z) + ...,\eqno(4.8)
$$

\noindent operator $\Delta $ acts on the transverse components of vector
$\vec {k}$: $\displaystyle\Delta = \frac{\partial
^2}{\partial n_x ^2} + \frac{\partial ^2}{\partial n_y ^2}$, where
$\vec {n} = \vec {k} / k$; $\displaystyle\overline {\,\theta ^2} =
N\int {\theta ^2} \frac{d\sigma _{coul} }{d\Omega }d\Omega $,
$\displaystyle\overline {\,\theta ^4} = N\int {\theta ^4}
\frac{d\sigma _{coul} }{d\Omega }d\Omega $, and etc.

If only the first term on the right-hand side of equation (4.8) is
taken into account, the integro-differential equation (4.6) is
reduced to a parabolic differential equation. The limiting angle
for such approximation is obtained from the condition [23]:

$$ \theta _{\max }^2 < \frac{16\Delta \hat {\rho }^{(0)}(\vec
{k},z)}{\Delta \Delta \hat {\rho }^{(0)}(\vec {k},z)}.\eqno(4.9)
$$
The solution of equation

$$\frac{d\hat {\rho }^{(0)}(\vec {k},z)}{dz} = \frac{\overline
{\,\theta ^2} }{4}\Delta \hat {\rho }^{(0)}(\vec {k},z)\eqno(4.10)
$$

\noindent for the initial condition $\hat {\rho }^{(0)}(\vec {k},z
= 0) = \hat {\rho }_0 \delta (n_x )\delta (n_y )$  and an infinite
medium is

$$\hat {\rho }^{(0)}(\vec {k},z) = \hat {\rho }_{0s} g_c(\vec{k}, z),\eqno(4.11) $$
where
\[
\hat {\rho }_{0s}=\frac{1}{3}I_0\hat{I}+\frac{1}{2}\vec{P}_0\vec{S}+\frac{1}{9}Q_{0ik}\hat{Q}_{ik}
\]
(see equation (3.3)), $\displaystyle g_c(\vec{k},
z)=\frac{1}{\pi\overline{\theta^2}z}e^{-\frac{(\vec{n}-\vec{n}_0)^2}{\overline{\theta^2}z}}$.
The unit vector $\vec {n}$ is counted from an arbitrary direction
$\vec {n}_0 $ (then direct the $z$-axis along $\vec {n}_0 )$,
$(\vec{n}-\vec{n}_0)^2=\theta^2$, where $\theta$ is the angle
between $\vec{k}$ and the $z$-axis directed along $\vec{n}_{0}$;
$\overline {\,\theta ^2}$ is the average squared angle of
scattering per unit path length. The average squared angle of
scattering at depth $z$ ($\overline {\,\theta ^2} z$) will further
be denoted by $\overline {\,\theta ^2_{z}}$, and its square root,
by ${\theta}_{z}$.

The applicability condition (4.9) of this solution  is equivalent
to the inequality $\theta \ll \theta_{z} $. It should also be
stated that for large $\theta $, the solution of equation (4.6)
will mainly  behave as  $ \sim \raise0.7ex\hbox{$1$}
\!\mathord{\left/{\vphantom {1 {\theta
^4}}}\right.\kern-\nulldelimiterspace}\!\lower0.7ex\hbox{${\theta^4}$}$
[24].
Thus, in the approximate equation (4.10), scattering at large
angles in a single scattering event is ignored. Indeed, for
$\theta \gg \theta_{z} $, the solution (4.11) decreases
exponentially, while the next term, corresponding to the solution
of the initial equation (4.6), decreases according to a power law
$\theta ^{ - 4}$. The angular distribution (4.11), therefore, does
not describe the characteristics of particles scattered at angles
$\theta \gg \theta_{z}$.

The expression for the density matrix in the zero-order
approximation (4.11) enables obtaining the correction  in the
first order perturbation theory $\hat {\rho }^{(1)}(\vec {k},z)$:
$$
\begin{array}{l}
 \displaystyle\hat {\rho }^{(1)}(\vec {k},z) = N\int_0^z {\int {G(\vec {k} - {\vec
{k}}'';z - z')} } \left( { - \sigma _{NC}^{\scriptscriptstyle 0}
\hat {\rho }^{(0)}({\vec {k}}'',z') - \frac{1}{2}\left( {\sigma
_{NC}^{\scriptscriptstyle\pm 1} - \sigma
_{NC}^{\scriptscriptstyle0} } \right)\left\{ {\hat {f}_s ({\vec
{k}}'',{\vec {k}}''),\hat {\rho
}^{(0)}({\vec {k}}'',z')} \right\} + } \right. \\[-10pt]
{}\\
 \displaystyle
 + \left. {\frac{2\pi i}{k}N \mathrm{Re} f_1 ^{nucl}(0)\left[ {\hat {f}_s ({\vec
{k}}'',{\vec {k}}''),\hat {\rho }^{(0)}({\vec {k}}'',z')} \right]}
\right)d\Omega _{\vec {k}''} dz'+ \\[-10pt]
{}\\
 \displaystyle
 + N\int_0^z {\int {G(\vec {k} - {\vec {k}}'';z - z' )} } \left( {\hat
{f}_{coul} ({\vec {k}}'',{\vec {k}}')\hat {\rho }^{(0)}({\vec
{k}}',z'
)\hat {f}^{\scriptscriptstyle + }_{nucl,coul} ({\vec {k}}',{\vec {k}}'') + } \right. \\[-10pt]
{}\\
 \displaystyle
 + \left. {\hat {f}_{nucl,coul} ({\vec {k}}'',{\vec {k}}')\hat {\rho
}^{(0)}({\vec {k}}',z' )\hat {f}^{\scriptscriptstyle + }_{coul}
({\vec {k}}',{\vec {k}}'')}
\right)d\Omega _{\vec {k}'} d\Omega _{\vec {k}''} dz' \\[-10pt]
{}\\
 \displaystyle
 + N\int_0^z {\int {G(\vec {k} - {\vec {k}}'';z - z' )} } \hat
{f}_{nucl,coul} ({\vec {k}}'',{\vec {k}}')\hat {\rho }^{(0)}({\vec
{k}}',z' )\hat {f}^{\scriptscriptstyle + }_{nucl,coul} ({\vec
{k}'},{\vec {k}''})d\Omega _{\vec {k}'} d\Omega _{\vec {k}''} dz'
,
 \end{array}\eqno(4.12)
$$

\noindent where $G(\vec {k} - {\vec {k}}'';z - z' )$ is the Green
function of equation (4.10): $\displaystyle G(\vec {k} - {\vec
{k}}'';z - z' ) = \frac{1}{\pi \overline {\theta ^2} |z - z'|}e^{
- \frac{(\vec {n} - {\vec {n}}'')^2}{\overline {\theta ^2} |z - z'
|}}.$ The amplitudes of Coulomb and nuclear scattering in the
general case are expressed by formula (2.1).

Thus, the solution of kinetic equation (4.1) allowing for nuclear
interaction has the following general form:
$$
\begin{array}{l}
 \displaystyle\hat {\rho }(\vec {k},z) \simeq \hat {\rho }^{(0)}(\vec
{k},z) + \hat {\rho }^{(1)}(\vec {k},z)\end{array},\eqno(4.13)
$$
where the main contribution  $\hat {\rho }^{(0)}(\vec {k},z)$ is
defined by expression (4.11), the correction in the first order
perturbation theory $\hat {\rho}^{(1)}(\vec {k},z)$ equals (4.12).

To represent the solution in the explicit form, we shall confine
ourselves to considering small scattering angles. In this case one
can obtain the following approximate expression for the amplitude
(2.1):

$$\hat {f}(\vec {k},{\vec {k}}') = A(\chi )\left[ {1 + b\chi (\vec
{S}\vec {\nu }) + c_1 \chi ^2\hat {Q}_{ik} \mu _i \mu _k + c_2 (1
- {\chi ^2} \mathord{\left/ {\vphantom {{\chi ^2} 4}} \right.
\kern-\nulldelimiterspace} 4)\hat {Q}_{ik} \mu _{1i} \mu _{1k} }
\right],\eqno(4.14)
$$

\noindent where the angular dependence of functions $b(\chi )$,
$c_1 (\chi )$, and  $c_2 (\chi )$ was supposed to be $b(\chi )
\sim \sin \chi $, $c_1 (\chi ) \sim \sin ^2\chi $, $c_2 (\chi )
\sim \cos ^2\chi \mathord{\left/ {\vphantom {\chi 2}} \right.
\kern-\nulldelimiterspace} 2$.

Integration of the right- and left-hand sides of the solution of
(4.13) over the final momentum $\vec {k}$ in the limits
corresponding to the variation of the axial angle from 0 to $2\pi
$  and the polar angle, from 0 to a certain $\vartheta _{det}$
gives the expression for the density matrix of the beam that has
passed through the area occupied by the detector with angular
width $\Delta \Omega $.

As has been stated above, for high energy deuterons, elastic scattering
occurs mainly at small angles. In this case in the solution of
(4.13) the general structure of the scattering amplitude in the
form (4.14) should be used for the nuclear amplitude. It can be
shown, however, that under the assumption of axial symmetry of the
collimator of the detector, the correction terms in the
polarization characteristics can be neglected with high accuracy.
These polarization characteristics correspond to the terms in
(4.14), which are proportional to $\chi $ and $\chi ^2$. This
assertion is also valid for such target thicknesses and deuteron
energies, for which the following inequalities are fulfilled:
$\overline {\theta ^2} z \ll 1$, $1 / kR_d \ll 1$, $1 / kR_c \ll
1$.

As a consequence, in the expression for the density matrix (4.13),
the amplitude $\hat {f}_{nucl,coul} (\vec {k},{\vec {k}}')$ of
scattering due to nuclear interaction is taken equal to:

$$\hat {f}_{nucl,coul} (\vec {k},{\vec {k}}') = d(\theta ) + d_1
(\theta )(\vec {S}\vec {n}_0 )^2,\eqno(4.15)
$$
$\vec {n}_0$ is the unit vector directed along the $z$-axis,
$d(\theta )$ is the spin-independent and $d_1 (\theta )$ is the
spin-dependent parts of the nuclear amplitude, $\theta$ is the
angle between   $\vec {k}$ and ${\vec {k}}'$.

For the Coulomb scattering amplitude $\hat {f}_{coul} (\vec
{k},{\vec {k}}')$, its spinless part $a(\theta )$ can be used with
high accuracy.

Integral characteristics of the deuteron beam are obtained by
substituting the explicit form of the density matrix (3.3) and the
scattering amplitude (4.15) into the solution (4.13). As a result,
we have:

$$
\begin{array}{l}
 \displaystyle{\cal I}(z) = (1 - e^{ - \frac{\vartheta _{det }^2 }{\overline {\,\theta ^2_{z}}
}})I_0 + \xi _1 I_0 + (\xi _2 + \xi _3 )\left[ {\frac{2}{3}I_0 +
\frac{1}{3}({\rm {\bf Q}}_0 \vec {n}_0 )\vec {n}_0 } \right],  \\[-10pt]
{}\\
{}\\
\displaystyle
 \vec {P}(z) = (1 - e^{ - \frac{\vartheta _{det }^2 }{\overline {\,\theta ^2_{z}}
}})\vec {P}_0 + \xi _1 \vec {P}_0 + \frac{1}{2}\xi _2 \left[ {\vec
{P}_0 + \vec {n}_0 (\vec {P}_0 \vec {n}_0 )} \right] + \xi _3 \vec
{n}_0 (\vec {P}_0 \vec {n}_0 )z - \frac{2}{3}\varphi _{eff} [\vec
{n}_0 \times ({\rm {\bf
Q}}_0 \vec {n}_0 )],  \\[-10pt]
{}\\
{}\\
\displaystyle
 {\cal Q}(z) = (1 - e^{ - \frac{\vartheta _{det }^2 }{\overline {\,\theta ^2_{z}}
}}){\rm {\bf Q}}_0 + \xi _1 {\rm {\bf Q}}_0 +  \\[-10pt]
{}\\
 \displaystyle
 + \xi _2 \left[ {{\rm {\bf Q}}_0 + \frac{1}{3}(3\vec {n}_0 \otimes \vec
{n}_0 - {\rm {\bf I}})I_0 - \frac{1}{2}\left( {({\rm {\bf Q}}_0
\vec {n}_0 ) \otimes \vec {n}_0 + \vec {n}_0 \otimes ({\rm {\bf
Q}}_0 \vec {n}_0 )} \right) + \frac{1}{3}{\rm {\bf I}}({\rm {\bf
Q}}_0 \vec {n}_0 )\vec {n}_0 }
\right] +  \\[-10pt]
{}\\
 \displaystyle
 + \xi _3 \left[ {\frac{1}{3}(3\vec {n}_0 \otimes \vec {n}_0 - {\rm {\bf
I}})I_0 - } \right.\frac{1}{2}\left( {({\rm {\bf Q}}_0 \vec {n}_0
) \otimes \vec {n}_0 + \vec {n}_0 \otimes ({\rm {\bf Q}}_0 \vec
{n}_0 )} \right) +
\frac{1}{2}\vec {n}_0 ^{\scriptscriptstyle\times} {\rm {\bf Q}}_0 \vec {n}_0 ^{\scriptscriptstyle\times} + \\[-10pt]
{}\\
 \displaystyle
 + \left. {\vec {n}_0 \otimes \vec {n}_0 \,\,({\rm {\bf Q}}_0 \vec {n}_0
)\vec {n}_0 - \frac{1}{6}{\rm {\bf I}}({\rm {\bf Q}}_0 \vec {n}_0
)\vec {n}_0 + \frac{1}{2}{\rm {\bf Q}}_0 } \right] -
\frac{3}{2}\varphi _{eff} \left( {[\vec {n}_0 \times \vec {P}_0 ]
\otimes \vec {n}_0 + \vec {n}_0 \otimes [\vec {n}_0 \times \vec
{P}_0 ]} \right).
 \end{array}\eqno(4.16)
$$

The parameters  $\xi _1 $, $\xi _2 $, and $\xi _3 $ are defined in terms of the relations:
$$
\begin{array}{l}
 \displaystyle\xi _1 (z) = - N\sigma _{NC}^{\scriptscriptstyle 0} \int\limits_0^z {\int{\int\limits_{\Delta
\Omega } {G(\vec {k} - {\vec {k}}'';z - z' )} } } g_{c} (\vec
{k}'',z') d\Omega_{\vec {k}''} d\Omega_{\vec {k}} dz' + \\[-10pt]
{}\\
 \displaystyle
 + N\int\limits_0^z {\int{\int{\int\limits_{\Delta \Omega } {G(\vec {k} -
{\vec {k}}'';z - z' )} } } } \left( {a({\vec {k}}'',{\vec
{k}}')d^{\scriptscriptstyle + }({\vec {k}}',{\vec {k}}'') +
d({\vec {k}}'',{\vec {k}}')a^{\scriptscriptstyle + }({\vec
{k}}',{\vec {k}}'')} \right)g_{c} \mbox{(}{\vec {k}}',z'
\mbox{)}d\Omega
_{\vec {k}''}d\Omega _{\vec {k}'} d\Omega _{\vec {k}} dz' + \\[-10pt]
{}\\
 \displaystyle
 + N\int\limits_0^z {\int {\int {\int\limits_{\Delta \Omega } {G(\vec {k} -
{\vec {k}}'';z - z' )d({\vec {k}}'',{\vec
{k}}')d^{\scriptscriptstyle + }({\vec {k}}',{\vec {k}}'')} } } }
g_{c} \mbox{(}{\vec {k}}',z' \mbox{)}d\Omega _{\vec {k}''}
d\Omega _{\vec {k}'} d\Omega _{\vec {k}} dz' , \\[-10pt]
{}\\
{}\\
\displaystyle
 \xi _2 (z) = - N(\sigma _{NC}^{\scriptscriptstyle\pm 1} - \sigma _{NC}^{\scriptscriptstyle0} )\int\limits_0^z
{\int {\int\limits_{\Delta \Omega } {G(\vec {k} - {\vec {k}}'';z -
z' )} } } g_{c} \mbox{(}{\vec {k}}'',z' \mbox{)}d\Omega _{\vec
{k}}'' d\Omega _{\vec{k}} dz' + \\[-10pt]
{}\\
 \displaystyle
 + N\int\limits_0^z {\int {\int {\int\limits_{\Delta \Omega } {G(\vec {k} -
{\vec {k}}'';z - z' )} } } } \left( {d_1 ({\vec {k}}'',{\vec
{k}}')a^{\scriptscriptstyle + }({\vec {k}}',{\vec {k}}'') +
a({\vec {k}}'',{\vec {k}}')d_1^{\scriptscriptstyle + }({\vec
{k}}',{\vec {k}}'')} \right)g_{c} \mbox{(}{\vec {k}}',z'
\mbox{)}d\Omega
_{\vec {k}''} d\Omega _{\vec {k}'} d\Omega _{\vec {k}} dz' + \\[-10pt]
{}\\
 \displaystyle
 + N\int\limits_0^z {\int {\int {\int\limits_{\Delta \Omega } {G(\vec {k} -
{\vec {k}}'';z - z' )} } } } \left( {d({\vec {k}}'',{\vec
{k}}')d_1^{\scriptscriptstyle + }({\vec {k}}',{\vec {k}}'') + d_1
({\vec {k}}'',{\vec {k}}')d^{\scriptscriptstyle + }({\vec
{k}}',{\vec {k}}'')} \right)g_{c} \mbox{(}{\vec {k}}',z'
\mbox{)}d\Omega
_{\vec {k}}'' d\Omega _{\vec {k}'} d\Omega _{\vec {k}} dz' , \\[-10pt]
{}\\
{}\\
\displaystyle
 \xi _3 (z) = N\int\limits_0^z {\int {\int {\int\limits_{\Delta \Omega }
{G(\vec {k} - {\vec {k}}'';z - z' )} } } } d_1 ({\vec {k}}'',{\vec
{k}}')d_1^{\scriptscriptstyle + }({\vec {k}}',{\vec {k}}'')g_{c}
\mbox{(}{\vec {k}}',z' \mbox{)}d\Omega _{\vec {k}''} d\Omega
_{\vec {k}'} d\Omega _{\vec {k}} dz' .
 \end{array}\eqno(4.17)
$$

The general structure of the solution  (4.16) is written in the
same way as the solution (3.4) for a thin target. For this reason,
all the conclusions derived in analyzing each of the terms
appearing  in the expression for the polarization
 vector in (3.4) are automatically (naturally) extended to the case of a thick target.

The rotation angle of the polarization vector with respect to vector $\vec {n}_0$ appears equal to

$$
\begin{array}{l}
 \displaystyle\varphi _{eff} = \frac{2\pi N}{k}\int\limits_0^z {\int {\int\limits_{\Delta
\Omega } {G(\vec {k} - {\vec {k}}'';z - z' )\left( {d_1 (\vec
{k},\vec {k}) + d_1^{\scriptscriptstyle + }(\vec {k},\vec {k})}
\right)} } } g_{c} \mbox{(}{\vec
{k}}'',z' \mbox{)}d\Omega _{\vec {k}''} d\Omega _{\vec {k}} dz' - \\[-10pt]
{}\\
 \displaystyle
 - \frac{iN}{2}\int\limits_0^z {\int {\int {\int\limits_{\Delta \Omega }
{G(\vec {k} - {\vec {k}}'';z - z' )} } } } \left( {d_1 ({\vec
{k}}'',{\vec {k}}')a^{\scriptscriptstyle + }({\vec {k}}',{\vec
{k}}'') - a({\vec {k}}'',{\vec {k}}')d_1^{\scriptscriptstyle +
}({\vec {k}}',{\vec {k}}'')} \right)g_{c} \mbox{(}{\vec {k}}',z'
\mbox{)}d\Omega _{\vec {k}''} d\Omega _{\vec {k}'} d\Omega _{\vec
{k}} dz' -
\\[-10pt]
{}\\
 \displaystyle
 - \frac{iN}{2}\int\limits_0^z {\int {\int {\int\limits_{\Delta \Omega }
{G(\vec {k} - {\vec {k}}'';z - z' )} } } } \left( {d_1 ({\vec
{k}}'',{\vec {k}}')d^{\scriptscriptstyle + }({\vec {k}}',{\vec
{k}}'') + d({\vec {k}}'',{\vec {k}}')d_1^{\scriptscriptstyle +
}({\vec {k}}',{\vec {k}}'')} \right)g_{c} \mbox{(}{\vec {k}}',z'
\mbox{)}d\Omega _{\vec {k}''} d\Omega _{\vec {k}'} d\Omega _{\vec
{k}} dz' .
 \end{array}\eqno(4.18)
$$

\subsection{Analysis of the problem of compensation of the
Coulomb-nuclear contributions in the expression for the rotation
angle of the polarization vector}

Let us consider in detail the contributions that are included in
the expression for the rotation angle of the polarization vector
of deuterons. The expression between the first brackets that
contain the nuclear amplitude $d_1 $ is the part of the rotation
angle, which is due to the coherent scattering of deuterons in
matter; the expression between the second brackets, which consists
of the product of $d_1 $ and the Coulomb amplitude $a$, describes
the effect of the Coulomb-nuclear interference; the last part in
$\varphi _{eff} $, which contains the product of the
spin-dependent and spinless parts of the nuclear amplitude, is the
correction to the deuteron spin rotation  from nuclear scattering.

It would be recalled, however, that nuclear amplitudes $d$ and
$d_1 $ are actually the scattering amplitudes modified by the
Coulomb interaction. The relation of $\hat {f}_{nucl,coul} (\vec
{k},{\vec {k}}')$ to "pure"\, Coulomb and "pure"\, nuclear
amplitudes is expressed by a general (2.9) or an approximate
(3.12) formula.

Substitution of $d_1 (\vec {k},{\vec {k}}')$ from (3.12) into the expression for $\varphi _{eff} $ (4.18) in the accepted approximation gives

$$
\begin{array}{l}
 \displaystyle\varphi _{eff} = \frac{2\pi N}{k} \mathrm{Re} d_1' (0)\int\limits_0^z {\int
{\int\limits_{\Delta \Omega } {G(\vec {k} - {\vec {k}}'';z - z' )}
} } g_{c} \mbox{(}{\vec {k}}'',z' \mbox{)}d\Omega _{\vec {k}''}
d\Omega _{\vec {k}}
dz' + \\[-10pt]
{}\\
 \displaystyle
 + \frac{iN}{4}\int\limits_0^z {\int {\int\limits_{\Delta \Omega } {G(\vec
{k} - {\vec {k}}'';z - z' )\left( {d_1 ' (\vec {k},{\vec
{k}}')a({\vec {k}}',\vec {k}) + a(\vec {k},{\vec {k}}')d_1 '
({\vec
{k}}',\vec {k})} \right. - } } } \\[-10pt]
{}\\
 \displaystyle
 \left. { - d_1 '^{*} (\vec {k},{\vec {k}}')a^{*}({\vec {k}}',\vec
{k}) - a^{*}(\vec {k},{\vec {k}}')d_1'^{*} ({\vec {k}}',\vec {k})}
\right)g_{c} \mbox{(}{\vec {k}}'',z' \mbox{)}d\Omega _{\vec
{k}''}'
d\Omega _{\vec {k}'} d\Omega _{\vec {k}} dz' - \\[-10pt]
{}\\
 \displaystyle
 - \frac{iN}{2}\int\limits_0^z {\int {\int {\int\limits_{\Delta \Omega }
{G(\vec {k} - {\vec {k}}'';z - z' )} } } } \left( {d_1 ' ({\vec
{k}}'',{\vec {k}}')a^{\scriptscriptstyle + }({\vec {k}}',{\vec
{k}}'') - a({\vec {k}}'',{\vec {k}}'){d}_{1}^{\scriptscriptstyle
+}({\vec {k}}',{\vec {k}}'')} \right)g_{c} \mbox{(}{\vec {k}}',z'
\mbox{)}d\Omega _{\vec {k}''} d\Omega _{\vec {k}'}' d\Omega _{\vec
{k}} dz' - \\[-10pt]
{}\\
 \displaystyle
 - \frac{iN}{2}\int\limits_0^z {\int {\int {\int\limits_{\Delta \Omega }
{G(\vec {k} - {\vec {k}}'';z - z' )} } } } \left( {d_1 ' ({\vec
{k}}'',{\vec {k}}'){d}'^{\scriptscriptstyle + }({\vec {k}}',{\vec
{k}}'') + {d}'({\vec {k}}'',{\vec
{k}}'){d}_{1}'^{\scriptscriptstyle +}({\vec {k}}',{\vec {k}}'')}
\right)g_{c} \mbox{(}{\vec {k}}',z' \mbox{)}d\Omega _{\vec {k}''}
d\Omega _{\vec {k}'} d\Omega _{\vec {k}} dz' .
 \end{array}\eqno(4.19)
$$

Expand the Green function $G(\vec {k} - {\vec {k}}';z -
z' )$ and the function $g_{c} \mbox{(}\vec {k},z' \mbox{)}$ into a series of orthogonal Legendre polynomials:
$$
\begin{array}{l}
 \displaystyle G(\vec {k} - {\vec {k}}';z - z' ) = \sum\limits_n {\left( {n +
\frac{1}{2}} \right)} G_n (z - z' )P_n (\cos \vartheta ), \\[-10pt]
{}\\
 \displaystyle
 g_{c} \mbox{(}{\vec {k}}',z\mbox{)} = \sum\limits_l {\left( {l + \frac{1}{2}}
\right)} g_{c\,l} \mbox{(}z\mbox{)}P_l (\cos {\vartheta }'), \\
 \end{array}\eqno(4.20)
$$
\noindent where $\vartheta$ is the angle between vectors $\vec
{k}$ and ${\vec {k}}'$; ${\vartheta }'$ is the angle between
vector ${\vec {k}}'$ and the $z$-axis.

In substitution of  relations (4.20) into (4.19), one should take
into account the condition of the orthogonality of the
polynomials:

$$ \int\limits_0^\pi
{P_j^m (\cos \vartheta )} P_r^m (\cos \vartheta )\sin \vartheta
d\vartheta = \frac{1}{j + 1 \mathord{\left/ {\vphantom {1 2}}
\right. \kern-\nulldelimiterspace} 2}\frac{(j + m)!}{(j -
m)!}\delta _{jr}\eqno(4.21)
$$
and  summation theorem:
$$ P_n (\cos \chi ) = P_n (\cos \vartheta )P_n (\cos \vartheta _1 )
+ 2\sum\limits_{m = 1}^n {\frac{(n - m)!}{(n + m)!}P_n^m (\cos
\vartheta )} P_n^m (\cos \vartheta _1 )\cos m(\varphi - \varphi _1
),\eqno(4.22)
$$

\noindent where $\chi $ is the angle between the two vectors that
make with the $z$-axis the angles $\vartheta $ and  $\vartheta
_1$, respectively.
Considering small deviation angles of particles from the initial
direction (the $z$-axis), let us take account of the fact that
large $n$ and  $l$ play the main part in the expansions (4.20). So
in view of of the below relation, one can pass  from the expansion
into a series of Legendre functions to that of Bessel functions
$$ \mathop {\lim }\limits_{n \to \infty } P_n (\cos \frac{\vartheta
}{k}) = J_0 (\vartheta ),\eqno(4.23)
$$
\noindent where $J_0 (\vartheta )$ is the zero-order Bessel
function. In this approximation the expansion coefficients
 $G_n (z - z' )$ and $g_{c\,l} \mbox{(}z\mbox{)}$ have a form:
$$
\begin{array}{l}
\displaystyle G_n (z - z' ) = \frac{1}{2\pi }e^{ -
\frac{\,n^2}{4}\overline {\theta ^2}
(z - z' )}, \\[-10pt]
{}\\
 \displaystyle
 g_{c\,l} \mbox{(}z\mbox{)} = \frac{1}{2\pi }e^{ - \frac{\,l^2}{4}\overline
{\theta ^2} z}.
 \end{array}\eqno(4.24)
$$

As a result,  expression (4.19) can be rewritten as follows:
$$
\begin{array}{l}
 \displaystyle\varphi _{eff} = \frac{2\pi N}{k} \mathrm{Re} d_1 ^\prime (0)\int\limits_{\Delta
\Omega } {g_{c} \mbox{(}\vec {k},z\mbox{)}} d\Omega _{\vec {k}} z + \\[-10pt]
{}\\
 \displaystyle
 + \frac{iN}{4}z\int {\int\limits_{\Delta \Omega } {\left( {d_1 '
(\vec {k},{\vec {k}}')a({\vec {k}}',\vec {k}) + a(\vec {k},{\vec
{k}}')d_1 ' ({\vec {k}}',\vec {k}) - d_1 '^{*} (\vec {k},{\vec
{k}}')a^\ast ({\vec {k}}',\vec {k}) - a^\ast (\vec {k},{\vec
{k}}')d_1 '^{*} ({\vec {k}}',\vec {k})} \right)} } g_{c}
\mbox{(}\vec
{k},z\mbox{)}d\Omega _{\vec {k}'} d\Omega _{\vec {k}} - \\[-10pt]
{}\\
 \displaystyle
 - \frac{iN}{2}z\int {\int\limits_{\Delta \Omega } {\left( {d_1 ^\prime
(\vec {k},{\vec {k}}')a^\ast (\vec {k},{\vec {k}}') - a(\vec
{k},{\vec {k}}'){d}_1'^{*}(\vec {k},{\vec {k}}')} \right)} } g_{c}
\mbox{(}{\vec
{k}}',z\mbox{)}d\Omega _{\vec {k}'} d\Omega _{\vec {k}} - \\[-10pt]
{}\\
 \displaystyle
 - \frac{iN}{2}z\int {\int\limits_{\Delta \Omega } {\left( {d_1 ^\prime
(\vec {k},{\vec {k}}'){d}'^{\scriptscriptstyle + }({\vec
{k}}',\vec {k}) + {d}'(\vec {k},{\vec
{k}}'){d}_1'^{\scriptscriptstyle + }({\vec {k}}',\vec {k})}
\right)} } g_{c} \mbox{(}{\vec {k}}',z\mbox{)}d\Omega _{\vec {k}'}
d\Omega _{\vec {k}} .
 \end{array}\eqno(4.25)
$$

Let now integration over the directions of the final momentum
$\vec {k}$   be made in the domain of the entire range of
variation in the scattering angle. In this case in (4.25), it is
possible to replace the integration variables $\vec {k}$ by ${\vec
{k}}'$ and vice versa.

For such deuteron energies
at which only Coulomb amplitude in the first Born approximation
can be taken into account, the Coulomb-nuclear contributions in
the expression for the rotation angle compensate one another.

As a
result, registration of the scattered and transmitted particles in
a $4\pi $ experimental geometry gives the following magnitude of
$\varphi _{eff} $:

$$ \varphi _{eff} = \frac{2\pi N}{k}z \mathrm{Re} \left[ {d_1 ' (0)
- \frac{ik}{2\pi }\int\!\!\!\int {d_1 ^\prime (\vec {k},{\vec
{k}}'){d}'^\ast (\vec {k},{\vec {k}}')g_{c} \mbox{(}{\vec
{k}}',z\mbox{)}d\Omega _{\vec {k}'} d\Omega _{\vec {k}} } }
\right].\eqno(4.26)
$$

In the case under consideration, the rotation angle is determined
by the sum of two terms: the coherent contribution, depending only
on a pure nuclear amplitude of scattering at zero angle and the
incoherent contribution due to single nuclear scattering.

In a real experiment, however, this case of a $4\pi$-geometry  is
generally not realized. That is why,  integration over $\vec {k}$
in (4.17) and  (4.18) or (4.19) should be made within  finite
limits.

\subsection{Calculation of the parameters of integral
characteristics of the beam.}

Integration over the angular variables in the parameters $\xi _1
$, $\xi _2 $, $\xi _3 $, and  $\varphi _{eff} $ is performed using
the expansion of the Green function $G(\vec {k} - {\vec {k}}';z -
z')$ and the function $g_{c}\mbox{(}\vec {k},z\mbox{)}$ into a
series of orthogonal Legendre polynomials according to formulas
(4.20)--(4.24).

In the explicit form, the parameters of the system (4.16) are as
follows:
$$
\begin{array}{l}
\displaystyle \xi _1 (z) = - N\sigma _{NC}^{\scriptscriptstyle 0}
(1 - e^{ - \frac{\vartheta _{det }^2 }{\overline {\theta
^2_{z}}}})z + 2\pi N z\int\limits_0^\infty {P(\chi ;\vartheta
_{\det } ,\overline {\,\theta ^2_{z}})\left( \frac{}{}{2
\mathrm{Re} [a(\chi )d^ * (\chi
)] + \vert d(\chi )\vert ^2} \right)} \chi d\chi , \\[-10pt]
{}\\
 \displaystyle
 \xi _2 (z) = - N(\sigma _{NC}^{\scriptscriptstyle\pm 1} - \sigma _{NC}^{\scriptscriptstyle 0} )(1 - e^{ -
\frac{\vartheta _{det }^2 }{\overline {\theta ^2_{z}}}})z + 2\pi
Nz\int\limits_0^\infty {P(\chi ;\vartheta _{\det } ,\overline
{\,\theta ^2_{z}})\left(\frac{}{}\mathrm{Re}[a(\chi )d_1^ * (\chi
)]\right.}+\\[-10pt]
{}\\
 \displaystyle+
 \left.{\mathrm{Re}[d(\chi
)d_1^
* (\chi )]}\frac{}{} \right)
\chi d\chi , \\[-10pt]
{}\\
 \displaystyle
 \xi _3 (z) = 2\pi Nz\int\limits_0^\infty {P(\chi ;\vartheta _{\det }
,\overline {\,\theta ^2_{z}})\vert d_1 (\chi )\vert ^2} \chi d\chi, \\
 \end{array}\eqno(4.27)
$$
$$\varphi _{eff} = \frac{2\pi N}{k}\mbox{Re}\left[ { {d_1
(0)\left( {1 - e^{ - \frac{\vartheta _{det }^2 }{\overline {\theta
^2_{z}}}}} \right)} - ik\int\limits_0^\infty {P(\chi ;\vartheta
_{\det } ,\overline {\,\theta ^2_{z}})\left(\frac{}{} {a^ * (\chi
)d_1 (\chi ) + d^ * (\chi )d_1 (\chi )} \right)\chi d\chi } }
\right]z,\eqno(4.28)
$$
\noindent where the integral $\displaystyle
P(\chi ;\vartheta _{det } ,\overline {\,\theta ^2_{z}}) \equiv
\int\limits_0^{\vartheta _{det } } {\int\limits_0^\infty
{\vartheta d\vartheta n\,dn\,e^{ - \frac{n^2}{4}\overline
{\,\theta ^2_{z}}}J_0 (n\vartheta )J_0 (n\chi )} } $ is
denoted in terms of the function $P(\chi ;\vartheta _{\det } ,\overline
{\,\theta ^2_{z}})$
The last expression is integrated over $n $.  As a result, the
function $P$ can be represented as the integral over $\vartheta $
of the expression containing a modified zero-order Bessel function
($I_0 (z) = J_0 (iz))$: $\displaystyle P(\chi ;\vartheta _{det }
,\overline {\,\theta ^2_{z}}) = \frac{2}{\overline {\,\theta
^2_{z}}}\int\limits_0^{\vartheta _{det } } {e^{ - \frac{\vartheta
^2 + \chi ^2}{\overline {\,\theta ^2_{z}}}}I_0 \left(
{\frac{2\vartheta \chi }{\overline {\,\theta ^2_{z}}}}
\right)\vartheta d\vartheta } $ .
The explicit form of the introduced function, which can be written
in a form of the following infinite series is $\displaystyle
P(\chi ;\vartheta _{\det } ,\overline {\,\theta ^2_{z}}) =
\sum\limits_{m = 0}^\infty {\frac{\Gamma (m + 1)}{m!\Gamma (m +
2)}( - 1)^m\left( {\frac{\vartheta _{det } ^2}{\overline {\,\theta
^2_{z}}}} \right)^{m + 1}\mbox{F}\left( { - m; - 1 -
m;1;\frac{\chi ^2}{\vartheta _{det } ^2}} \right)} ,$ where
$\mbox{F} $ is the gaussian hypergeometric function.

\subsection{Analysis of the contributions to the rotation angle $\varphi _{eff}$ for the case of scattering by a thick target}

Rewrite expression (4.28) for two limiting values of the  angle
$\vartheta _{det }$ of the collimator of the detector: $\vartheta
_{det } \ll \theta_z $ and  $\vartheta _{det }\gg\theta_z$.

Before considering the angles $\vartheta _{\det } \gg \theta_z $,
let us indicate the following points. As has been stated above,
the solution (4.11), which describes the angular distribution of
the deuterons due to  Coulomb interaction holds true for the
scattering angles $\theta \ll \theta_z $.
Since we analyze the  integral characteristics of the beam, the
solution (4.11) can also be used for $\vartheta _{\det } \gg
\theta_z $. The stated approximation means that the contribution
of singly scattered particles due to Coulomb interaction is
neglected. Moreover, large values of $\theta $  make  zero
contribution to the integral
$\displaystyle\int\limits_0^{\vartheta _{det } } {\rho ^{(0)}(\vec
{k},z)d\Omega } $ and a small contribution to
$\displaystyle\int\limits_0^{\vartheta _{det } } {\rho ^{(1)}(\vec
{k},z)d\Omega } $ (the major contribution to the integral comes
from small  $\theta$ ).

In the first case, in integration of the second term in (4.8), it
can be assumed that  $J_0 (n\vartheta ) \approx 1$. Let $\varphi
_{eff}^{\scriptscriptstyle{\mathrm{I}}} $ denote the rotation
angle of the beam that has passed through the area occupied by the
detector with angular width $\vartheta _{\det } \ll \theta_z $.
The explicit expression for $\varphi
_{eff}^{\scriptscriptstyle{\mathrm{I}}} $ has a form:

$$ \varphi _{eff}^{\scriptscriptstyle{\mathrm{I}}} = \frac{2\pi
N}{k}\frac{\vartheta _{det }^2 }{\overline {\,\theta ^2_{z}}
}\,\mbox{Re}\left[ {d_1 (0) - ik\int\limits_0^\infty {\chi d\chi }
e^{ - \frac{\chi ^2}{\overline {\,\theta ^2_{z}}}}\left(\frac{}{}
{a^
* (\chi )d_1 (\chi ) + d^ * (\chi )d_1 (\chi )}\frac{}{} \right)}
\right]z.\eqno(4.29)
$$

For another limiting case $\vartheta _{det } \gg\theta_{z}$, relation  (4.28) is written as:

$$\varphi _{eff}^{\scriptscriptstyle{\mathrm{II}}} \simeq \frac{2\pi N}{k}\mbox{Re}\left[ {d_1
(0) - ik\int\limits_0^\infty {\chi d\chi } \left( \frac{}{}{a^ *
(\chi )d_1 (\chi ) + d^ * (\chi )d_1 (\chi )} \frac{}{}\right)}
\right]z.\eqno(4.30)
$$
One can easily see that in the case when the whole beam gets into
the detector, the expression for the rotation angle $\varphi
_{eff}^{\scriptscriptstyle{\mathrm{II}}} $ is similar to that for
$\varphi _{eff} $, which was obtained in describing the passage of
the deuteron beam through a very thin target in the case of a wide
experimental geometry (3.8).

As it has been done in the previous section, let us represent
$\varphi _{eff} $ as a sum of three terms:

$$ \varphi _{eff} (\vartheta _{det} ) = \varphi _0 (\vartheta
_{det} ) + \varphi _{nc} (\vartheta _{det} ) + \varphi _{nn}
(\vartheta _{det} ),\eqno(4.31)
$$
\noindent where $\displaystyle\displaystyle\varphi _0 (\vartheta
_{det} ) = \frac{2\pi N}{k}\mbox{Re}\,d_1 (0)\left(1 - e^{ -
{\vartheta _{det }^2 } \mathord{\left/ {\vphantom {{\vartheta
_{det }^2 } {\overline {\,\theta ^2_{z}}}}} \right.
\kern-\nulldelimiterspace} {\overline {\theta ^2_{z}}}}\right)z$
is the contribution of the coherent scattering to the rotation of the polarization vector;
$\displaystyle\varphi _{nc} (\vartheta _{det} ) = 2\pi N
\mathrm{Im}\left[ {\int\limits_0^\infty {P(\chi ;\vartheta _{det}
,\overline {\,\theta ^2_{z}})a^
* (\chi )d_1 (\chi )\chi d\chi } } \right]z$ is the part of the rotation angle, which describes the
effect of the Coulomb-nuclear interference; $\varphi _{nn}
(\vartheta _{det} ) =\displaystyle 2\pi N \mathrm{Im}\left[
{\int\limits_0^\infty {P(\chi ;\vartheta _{det} ,\overline
{\,\theta ^2_{z}})d^ * (\chi )d_1 (\chi )\chi d\chi } } \right]z$
is the correction to the rotation of the deuteron spin from a pure
nuclear scattering.
Obtain the numerical values of these contributions for the energy
of 500 MeV, $k = 0.74 \cdot 10^{14}$cm$^{ - 1}$. Then we have
$kR_d = 31.82$ and the value of the average squared angle of
deuteron scattering by a carbon target per unit length
 $\displaystyle\overline {\theta ^2} = 16\pi
NZ^2\left( {\frac{e^2}{pv}} \right)^2\mbox{ln}(137\, Z ^{-1/3})$
equals $0.23 \cdot 10^{ - 4}$cm$^{ - 1}$. The maximum value of the
rotation angle $\varphi _0 $ is obtained at $\vartheta _{det } \gg
\theta_{z}$:\, $\varphi _0 \simeq 0.12 \cdot 10^{ - 3}z$.

Consider the relation ${\varphi _{nc} } \mathord{\left/ {\vphantom
{{\varphi _{nc} } {\varphi _0 }}} \right.
\kern-\nulldelimiterspace} {\varphi _0 }$.
Calculate the explicit form of the Coulomb-nuclear correction, using the expression for the Coulomb amplitude
corresponding to scattering by a screened Coulomb potential in the first Born approximation:

$$a(\theta ) = - 2\frac{mZe^2}{\hbar ^2}R_c^2 \frac{1}{1 + k^2R_c^2
\theta ^2}.\eqno(4.32)
$$

\noindent  In the case of scattering by carbon,  $kR_{coul}= 2 \cdot
10^5$. Using (3.7) and (4.32), one can obtain for the whole range of scattering angles:

$$ \frac{\varphi _{nc} }{\varphi _0 } = - 2\frac{\mathrm{Im} d_1 }{\mathrm{Re} d_1
}\frac{mZe^2}{\hbar ^2k}\frac{k^2R_{coul}^2 }{1 - e^{ - {\vartheta
_{det }^2 } \mathord{\left/ {\vphantom {{\vartheta _{det }^2 }
{\overline {\,\theta ^2_{z}}}}} \right. \kern-\nulldelimiterspace}
{\overline {\,\theta ^2_{z}}}}}\int\limits_0^\infty {P(\chi
;\vartheta _{det} ,\overline {\,\theta ^2_{z}})\frac{1}{1 +
k^2R_c^2 \chi ^2}e^{ - \frac{\chi ^2k^2R_d^2 }{4}}\chi d\chi
}\eqno(4.33)
$$
Plot this dependence for three values of $\overline {\,\theta
^2_{z}}$ (since the value of energy is fixed, this corresponds to
three values of the target thickness $z$):

\begin{figure}[htbp]
\centerline{\includegraphics[width=3.87in,height=2.55in]{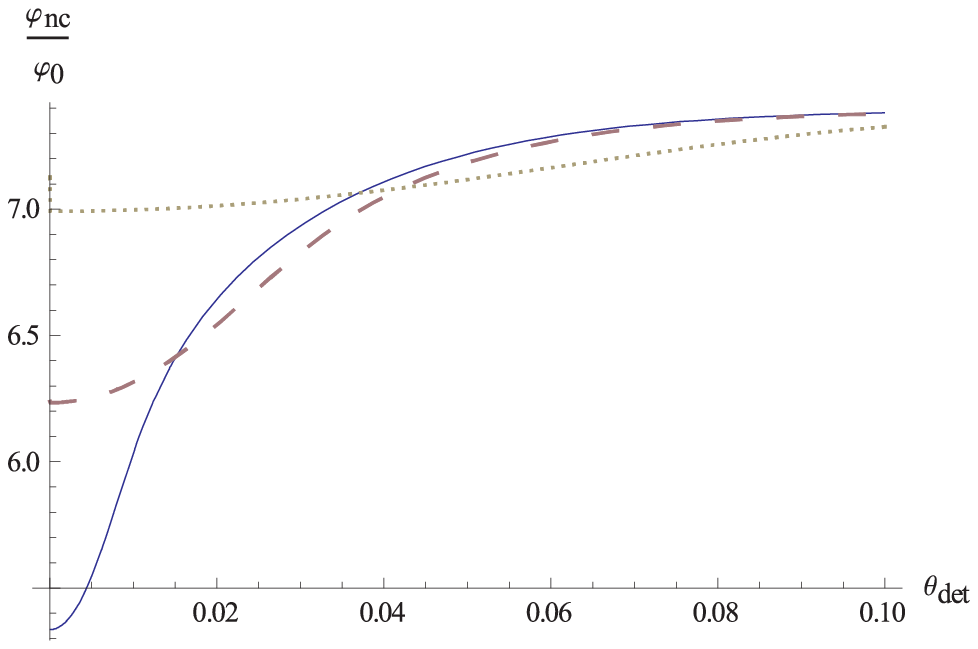}}
\label{fig3}
\end{figure}

$\\[-2pt]
{}\\$ {Fig. 3: Relative contribution of the Coulomb-nuclear
interference as a function of the angle of the detector for
$\overline {\,\theta ^2_{z}}_{1}=0.23\cdot10^{-4}$ cm$^{-1}$
(solid curve), $\overline {\,\theta ^2_{z}}_{2}=0.23\cdot10^{-3}$
cm$^{-1}$ (dashed curve) and $\overline {\,\theta
^2_{z}}_{3}=0.23\cdot10^{-2}$ cm$^{-1}$ (dotted curve).}

\bigskip

A  rather simple analytical expression for ${\varphi _{nc} }
\mathord{\left/ {\vphantom {{\varphi _{nc} } {\varphi _0 }}}
\right. \kern-\nulldelimiterspace} {\varphi _0 }$ can be obtained for two limiting values of $\vartheta _{\det } $.

Let $\vartheta _{\det } \ll \theta_{z} $. From (4.29) we have that
for small values of the  angles of the detector, the relation
${\varphi _{nc} } \mathord{\left/ {\vphantom {{\varphi _{nc} }
{\varphi _0 }}} \right. \kern-\nulldelimiterspace} {\varphi _0 }$
is the function independent of $\vartheta _{det } $:

$$\left( {\frac{\varphi _{nc} }{\varphi _0 }}
\right)^{\scriptscriptstyle{\mathrm{I}}} = \frac{\mathrm{Im} d_1
}{\mathrm{Re} d_1 }\frac{mZe^2}{\hbar ^2k}\mbox{Ei}\left[ {\mbox{
- }\frac{\mbox{1}}{k^2R_c^2 }\left( {\frac{1}{\overline {\,\theta
^2_{z}}} + \frac{k^2R_d^2 }{4}} \right)} \right]\exp \left[
{\mbox{ - }\frac{\mbox{1}}{k^2R_c^2 }\left( {\frac{1}{\overline
{\,\theta ^2_{z}}} + \frac{k^2R_d^2 }{4}} \right)}
\right].\eqno(4.34)
$$

When $\vartheta _{\det } \to \infty $ the expression ${\varphi
_{nc} } \mathord{\left/ {\vphantom {{\varphi _{nc} } {\varphi _0
}}} \right. \kern-\nulldelimiterspace} {\varphi _0 }$ tends to its limiting value (see Fig. 1):

$$ \left( {\frac{\varphi _{nc} }{\varphi _0 }}
\right)^{\scriptscriptstyle{\mathrm{II}}} = \frac{\mathrm{Im} d_1
}{\mathrm{Re} d_1 }\frac{mZe^2}{\hbar ^2k}\mbox{Ei}\left[ {\mbox{
- }\frac{R_d^2 }{4R_c^2 }} \right].\eqno(4.35)
$$

According to (4.34) for such energies and target thicknesses when
$\theta_{z}\gg\theta_{n}$, the relation $\left( {{\varphi _{nc} }
\mathord{\left/ {\vphantom {{\varphi _{nc} } {\varphi _0 }}}
\right. \kern-\nulldelimiterspace} {\varphi _0 }}
\right)^{\scriptscriptstyle{\mathrm{I}}}$ tends to $\left(
{{\varphi _{nc} } \mathord{\left/ {\vphantom {{\varphi _{nc} }
{\varphi _0 }}} \right. \kern-\nulldelimiterspace} {\varphi _0 }}
\right)^{\scriptscriptstyle{\mathrm{II}}}$, i.e., weakly depends
on $\vartheta _{det } $. If the parameter
$\theta_{z}\ll\theta_{n}$, then for $\vartheta _{det } \to 0$ the
relation  $\left( {{\varphi _{nc} } \mathord{\left/ {\vphantom
{{\varphi _{nc} } {\varphi _0 }}} \right.
\kern-\nulldelimiterspace} {\varphi _0 }}
\right)^{\scriptscriptstyle{\mathrm{I}}}$ decreases with
decreasing target thickness. The corresponding contribution can be
neglected in the case when $\displaystyle\overline {\,\theta
^2_{z}} \ll \theta_{n}^{2}\exp \left( {\mbox{C} - \frac{\hbar
^2k}{mZe^2}\frac{\mathrm{Re} d_1 }{\mathrm{Im} d_1 }} \right)$,
where $\mbox{C}$ is the Euler–Mascheroni constant, $\mbox{C
=0.5772}$. For the energy of 500 MeV, the value of $z$ that
satisfies this condition is $z \ll 10^{ - 6}$ cm. When
$\displaystyle \overline {\,\theta ^2_{z}} \gtrsim
\theta_{n}^{2}\exp \left( {\mbox{C} - \frac{\hbar
^2k}{mZe^2}\frac{\mathrm{Re} d_1 }{\mathrm{Im}d_1 }} \right)$, the
value of $\varphi _{nc} $ at $\vartheta _{det } \ll \theta_{z} $
is of the order of or greater than the main contribution $\varphi
_0 $.

With increasing $\vartheta _{det } $, the relation ${\varphi _{nc}
} \mathord{\left/ {\vphantom {{\varphi _{nc} } {\varphi _0 }}}
\right. \kern-\nulldelimiterspace} {\varphi _0 }$ grows,
approaching $\displaystyle\left( {{\varphi _{nc} } \mathord{\left/
{\vphantom {{\varphi _{nc} } {\varphi _0 }}} \right.
\kern-\nulldelimiterspace} {\varphi _0 }}
\right)^{\scriptscriptstyle{\mathrm{I}}} = \frac{\mathrm{Im} d_1
}{\mathrm{Re}d_1 }\frac{mZe^2}{\hbar ^2k}\left[ {\mbox{C +
ln}\left( {\frac{R_d^2 }{4R_c^2 }} \right)} \right]$, and this
value of the contribution does not depend on the target thickness.
From this follows that when $\vartheta _{\det } \gg \theta_{z}$,
the contribution of the Coulomb-nuclear scattering cannot be
neglected for any $z$: $\varphi_{nc} $ is one order of magnitude
larger than the main contribution $\varphi _0 $.

\subsection{Rotation angle including the Coulomb contribution to
the spin part of nuclear amplitude $d_{1}(0)$}

Let us consider in more detail the expression for the rotation angle including the Coulomb contribution to the amplitude $d_1 (0)$.
For the analysis of the total contribution of the Coulomb-nuclear interference, it is convenient to write equation (4.31) in a form:

$$\varphi _{eff} = {\varphi }'_0 + \varphi _{nc}^{tot} + \varphi
_{nn} ,\eqno(4.36)
$$

\noindent where $\varphi _{nc}^{tot}  \quad $ is the total
contribution of the Coulomb-nuclear interference into the rotation
angle of the polarization vector. Within the limit of small
scattering angles,  its explicit form reads: :
$\displaystyle\varphi _{nc}^{tot} = - \frac{2\pi
N}{k}\frac{mZe^2}{\hbar ^2k^2}\mathrm{Im}d_1 \left\{
{2k^2R_{coul}^2 \int\limits_0^\infty {f(\chi ;\vartheta _{\det }
,\overline {\,\theta ^2_{z}})\frac{1}{1 + k^2R_c^2 \chi ^2}e^{ -
\frac{\chi ^2k^2R_d^2 }{4}}\chi d\chi } + \mbox{Ei}\left( { -
\frac{R_d^2 }{4R_c^2 }}
\right)\left(1-e^{-\vartheta^{2}_{det}/\overline {\,\theta
^2_{z}}}\right)}\right\}z.$ Basing on the above considerations,
one may immediately see that when the condition $\theta_{z}\gg
\theta_{n}$ is fulfilled, the total Coulomb-nuclear contribution
vanishes.

Let us analyze the behavior of the Coulomb-nuclear interference when $\theta_{z}\ll \theta_{n}$.
With this aim in view, consider the relative contribution of $\varphi _{nc}^{tot} $:

$$\begin{array}{l}\displaystyle\frac{\varphi _{nc}^{tot} }{{\varphi }'_0 } = -
\frac{\mathrm{Im}d_1' }{\mathrm{Re}{d}'_1 }\frac{mZe^2}{\hbar
^2k}\frac{1}{1 - e^{ - {\vartheta _{\det }^2 } \mathord{\left/
{\vphantom {{\vartheta _{\det }^2 } {\overline {\,\theta
^2_{z}}}}} \right. \kern-\nulldelimiterspace} {\overline {\,\theta
^2_{z}}}}}\left\{2k^2R_{coul}^2 \int\limits_0^\infty {f(\chi
;\vartheta _{\det } ,\overline {\,\theta ^2_{z}})\frac{1}{1 +
k^2R_c^2 \chi ^2}e^{ - \frac{\chi
^2k^2R_d^2 }{4}}\chi d\chi }+\right.\\[-10pt]
{}\\
 \displaystyle+\left.\mbox{Ei}\left( {
- \frac{R_d^2 }{4R_c^2 }}
\right)\left(1-e^{-\vartheta^{2}_{det}/\overline {\,\theta
^2_{z}}}\right)\right\} ,
\end{array}\eqno(4.37)$$

\noindent where $\mathrm{Re}{d}'_1 $ is the spin-dependent part of a pure nuclear amplitude of scattering.
According to (3.12), the relation between the amplitudes $\mathrm{Re}{d}'_1 $ and  $\mathrm{Re}{d}_1 $ is determined by formula
$$ \mathrm{Re}{d}'_1 = \mathrm{Re}d_1 + \mathrm{Im}d_1 \frac{mZe^2}{\hbar ^2k}\mbox{Ei}\left[
{\mbox{ - }\frac{R_d^2 }{4R_c^2 }} \right].\eqno(4.38)
$$
The calculated value is
 ${\mathrm{Re}d_1 } \mathord{\left/ {\vphantom
{{\mathrm{Re}d_1 } {\mathrm{Re}{d}_1 }}} \right.
\kern-\nulldelimiterspace} {\mathrm{Re}{d}'_1 } = 0.12$.
Using this estimate, let us plot the dependence ${\varphi
_{nc}^{tot} } \mathord{\left/ {\vphantom {{\varphi _{nc}^{tot} }
{{\varphi }'_0 }}} \right. \kern-\nulldelimiterspace} {{\varphi
}'_0 }$ for three values of the average squared angle $\overline
{\,\theta ^2_{z}}$ of multiple scattering at depth $z$:

\begin{figure}[htbp]
\centerline{\includegraphics[width=3.87in,height=2.62in]{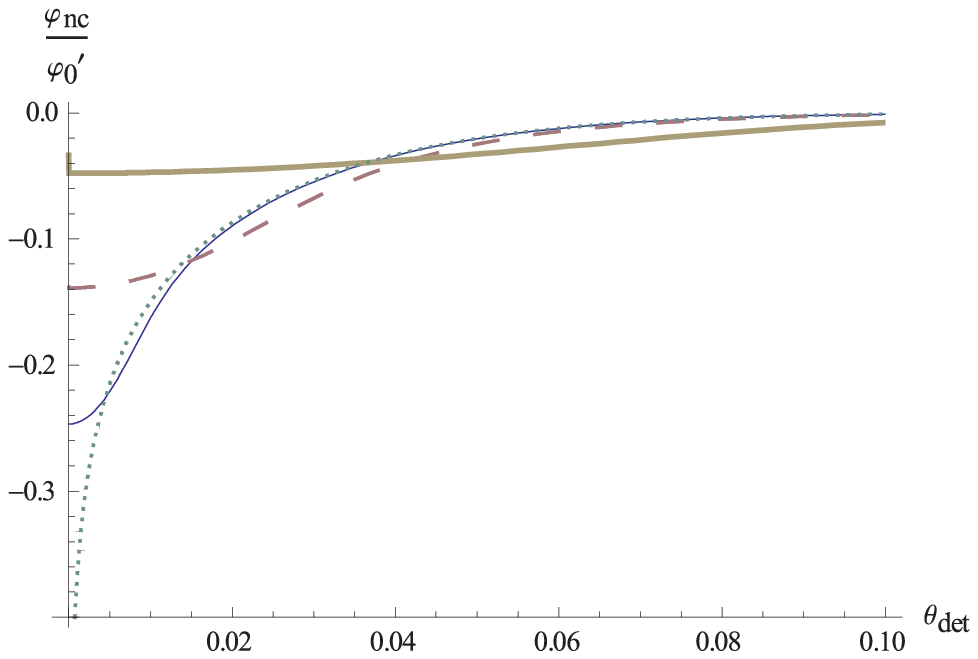}}
\label{fig2}
\end{figure}

$\\[-2pt]
{}\\${Fig. 4: Relative contribution of the Coulomb-nuclear
interference ${\varphi _{nc}^{tot} } \mathord{\left/ {\vphantom
{{\varphi _{nc}^{tot} } {\varphi _0 '}}} \right.
\kern-\nulldelimiterspace} {\varphi _0 '}$ as a function of the
angle of the detector for $\overline {\,\theta
^2_{z}}_{1}=0.23\cdot10^{-4}$ cm$^{-1}$ (solid curve), $\overline
{\,\theta ^2_{z}}_{2}=0.23\cdot10^{-3}$ cm$^{-1}$ (dashed curve)
and $\overline {\,\theta ^2_{z}}_{3}=0.23\cdot10^{-2}$ cm$^{-1}$
(thick solid curve).}

\bigskip

It is seen in Fig. 4 that the maximum  absolute value of $\vert
\varphi _{nc}^{tot} \vert $ is achieved at the angles of the
detector $\vartheta _{\det } \ll \theta_{z}$:

$$\begin{array}{l}\displaystyle\left( {\frac{\varphi _{nc}^{tot} }{{\varphi }'_0 }}
\right)_{\max } = \frac{\mathrm{Im}d_1 }{\mathrm{Re}{d}'_1
}\frac{mZe^2}{\hbar ^2k}\left\{ {\mbox{Ei}\left[ {\mbox{ -
}\frac{\mbox{1}}{k^2R_c^2 }\left( {\frac{1}{\overline {\,\theta
^2_{z}}} + \frac{k^2R_d^2 }{4}} \right)} \right]\exp \left[
{\mbox{ - }\frac{\mbox{1}}{k^2R_c^2 }\left( {\frac{1}{\overline
{\,\theta ^2_{z}}} + \frac{k^2R_d^2 }{4}} \right)} \right]}\right. -\\[-10pt]
{}\\
 \displaystyle\left.
\mbox{Ei}\left[ {\mbox{ - }\frac{R_d^2 }{4R_c^2 }} \right]
\right\}.
\end{array}\eqno(4.39)$$

For considered energy, at target thickness $z \le 0.01$ cm, the
magnitude of $\displaystyle\vert \varphi _{nc}^{tot} \vert $  is
of the order of ${\varphi }'_0 $. With further decrease in $z$,
the contribution of $\vert \varphi _{nc}^{tot} \vert $ grows and
reaches its maximum value $\displaystyle\frac{\mathrm{Im}d_1
}{\mathrm{Re}{d}'_1 }\frac{mZe^2}{\hbar ^2k}\mbox{Ei}\left[
{\mbox{ - }\frac{R_d^2 }{4R_c^2 }} \right]$ at $z \to 0$. It is
easy to see that it is exactly equal to the maximum absolute value
of the relative value of the total Coulomb-nuclear contribution
$\left| {{\varphi _{nc}^{tot} } \mathord{\left/ {\vphantom
{{\varphi _{nc}^{tot} } {{\varphi }'_0 }}} \right.
\kern-\nulldelimiterspace} {{\varphi }'_0 }} \right|$ for the case
of a thin target (3.16).
The maximum value of this quantity is also achieved for small
$\vartheta _{det } $, namely,  within the  limit $\vartheta _{det
} \to 0$. Moreover, for a thin target,  the dependence of the
relation ${\varphi _{nc}^{tot}} \mathord{\left/ {\vphantom
{{\varphi _{nc}^{tot} } {{\varphi }'_0 }}} \right.
\kern-\nulldelimiterspace} {{\varphi }'_0 }$ on the detector angle
$\vartheta _{det } $ is, in fact, the limit to which the
Coulomb-nuclear contribution tends when  deuterons are scattered
by a thick target, if the parameter $\overline {\,\theta ^2_{z}}
\to 0$ (dotted line in Fig. 4).

Indeed, at $\overline
{\,\theta ^2_{z}} \to 0$, the introduced function $f(\chi
;\vartheta _{\det } ,\overline {\,\theta ^2_{z}})$ becomes equal to  $f(\chi ;\vartheta
_{det } ,0) = \int\limits_0^{\vartheta _{det } } {\delta
(\vartheta - \chi )d\vartheta } $.
As a result, for the rotation angle $\varphi _{eff} $ defined by
formula  (4.28), we obtain the expression for $\varphi _{eff} $ in
the case of deuteron scattering by a thin target (3.8). It is
clear because within the limit $\overline {\,\theta ^2_{z}} \to
0$, the contribution of multiple scattering can be neglected for
given energies and (or) target thicknesses. This statement is true
even for $\theta_{z} \ll \theta_{c}$, i.e., for the case when the
value of the average squared angle of multiple scattering is much
smaller than the diffraction angle for Coulomb scattering. In this
case, similarly to the case of scattering by a thin target, one
can introduce the detector angle $\vartheta _{det
}^{\scriptscriptstyle comp} $ at which one may consider that the
interference contributions compensate each other.
Thus, for $\vartheta _{det } \ll \theta _n $, the contribution  of $\vert \varphi _{nc}^{tot} \vert $ is
comparable in magnitude to ${\varphi }'_0 $, i.e., the Coulomb-nuclear terms are not compensated.
When $\vartheta _{det } > \theta _n $, the total contribution of the Coulomb-nuclear interference can be neglected.

On the other hand, when $\theta_{z} \gg \theta_{c}$, the
dependence of ${\varphi _{nc}^{tot} } \mathord{\left/ {\vphantom
{{\varphi _{nc}^{tot} } {{\varphi }'_0 }}} \right.
\kern-\nulldelimiterspace} {{\varphi }'_0 }$ on $\vartheta _{det }
$ becomes more smooth (Fig. 4).
In this case the the maximum absolute value of   $\left| {{\varphi _{nc}^{tot} }
\mathord{\left/ {\vphantom {{\varphi _{nc}^{tot} } {{\varphi }'_0
}}} \right. \kern-\nulldelimiterspace} {{\varphi }'_0 }} \right|$
diminishes with growing parameter $\overline {\,\theta ^2_{z}}$:
$$ \left( {\frac{\varphi _{nc}^{tot} }{{\varphi }'_0 }}
\right)_{\max } = \frac{\mathrm{Im}d_1' }{\mathrm{Re}{d}'_1
}\frac{mZe^2}{\hbar ^2k}\ln \left[ {\frac{\mbox{4}}{k^2R_d^2
}\left( {\frac{1}{\overline {\,\theta ^2_{z}}} + \frac{k^2R_d^2
}{4}} \right)} \right].\eqno(4.40)
$$
From this follows that at the energy of 500 MeV and the target thickness $z \gg
1$ cm, up to the accuracy of 10{\%}, one may consider that the Coulomb-nuclear interference due to
incoherent scattering compensates the Coulomb-nuclear contribution to the amplitude $d_1 (0)$ for
all values of $\vartheta _{det } $. The expression for the rotation angle $\varphi_{eff} $ will also take quite a
simple form (below, it is shown that the nuclear part of  $\varphi _{nn} $ can as well be neglected in comparison with ${\varphi }'_0 $):

$$\varphi _{eff} \simeq {\varphi }'_0 = \frac{2\pi
N}{k}\mbox{Re}{d}'_1 (1 - e^{ - \frac{\vartheta _{det }^2
}{\overline {\theta ^2} z}})z.\eqno(4.41)
$$

Using (4.38), we obtain that the maximum value of this quantity is
${\varphi }'_0 \simeq 10^{ - 2}z$, i.e., one order of magnitude
larger than $\varphi _0 $.

Thus, for the energies in the region  $0.1\div 1$ GeV, one can indicate two main domains of
variability of the value of the  multiple scattering parameter $\overline {\,\theta ^2_{z}}$:

1.  \,\,$ \theta_{z}\ll \theta_{c}$. Here the dependence of the
function ${\varphi _{nc}^{tot} } \mathord{\left/ {\vphantom
{{\varphi _{nc}^{tot} } {{\varphi }'_0 }}} \right.
\kern-\nulldelimiterspace} {{\varphi }'_0 }$ on the angle
$\vartheta _{det }$ of the detector is close to the similar
dependence in the case of  a thin target.

2. \,\,$\theta_{z}\gtrsim \theta_{c}$. In this range of energies
and target thicknesses the dependence of the magnitude of the
relative  Coulomb-nuclear contribution on  $\vartheta _{det }$ is
quite smooth. The total contribution of the Coulomb-nuclear
interference can be neglected with quite a good accuracy.

\subsection{Nuclear contribution to the rotation angle}

The relative contribution of nuclear scattering of deuterons by target nuclei
${\varphi _{nn} } \mathord{\left/ {\vphantom {{\varphi _{nn} }
{\varphi _0 }}} \right. \kern-\nulldelimiterspace} {\varphi _0 }$
is obtained, using the approximate expression (3.7):

$$\frac{\varphi _{nn} }{\varphi _0 } = \frac{1}{kR_d^2 }\left(
{\mathrm{Re}d\frac{\mathrm{Im}d_1 }{\mathrm{Re}d_1 } -
\mathrm{Im}d} \right)\left[ {1 - \exp \left( { - \frac{k^2R_d^2
\vartheta _{\det }^2 }{2 + k^2R_d^2 \,\overline {\,\theta
^2_{z}}}} \right)} \right]\frac{1}{1 - e^{ - {\vartheta _{\det }^2
} \mathord{\left/ {\vphantom {{\vartheta _{\det }^2 } {\overline
{\theta ^2} z}}} \right. \kern-\nulldelimiterspace} {\overline
{\,\theta ^2_{z}}}}}.\eqno(4.42)
$$

Similarly to the previous case, let us represent the dependence of
${\varphi _{nn} } \mathord{\left/ {\vphantom {{\varphi _{nn} }
{\varphi _0 }}} \right. \kern-\nulldelimiterspace} {\varphi _0 }$
on the angle of the detector for three values of average squared
angle of multiple scattering at depth $z$:

\bigskip

\begin{figure}[htbp]
\centerline{\includegraphics[width=3.87in,height=2.46in]{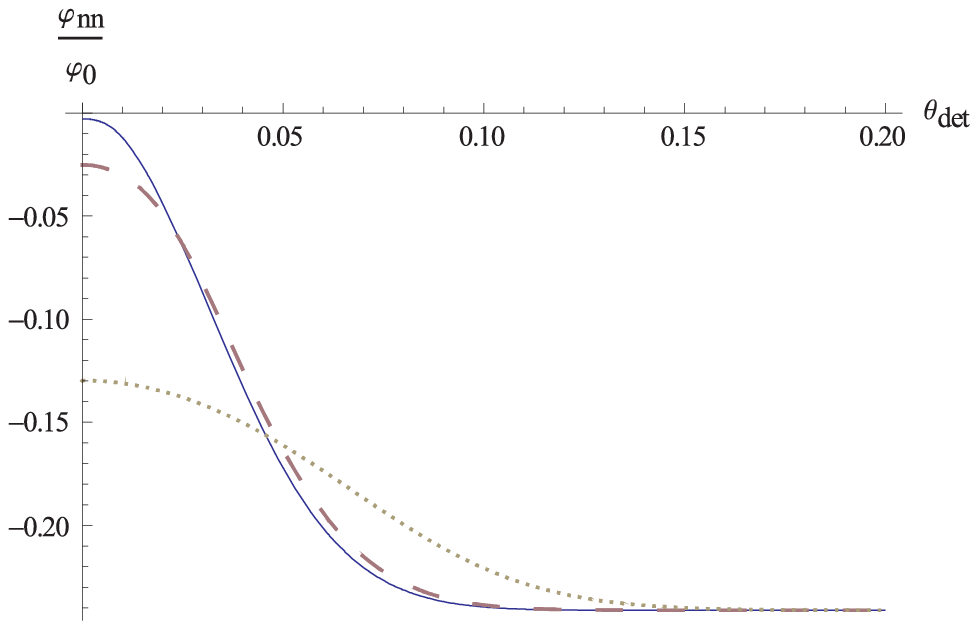}}
\label{fig3}
\end{figure}

$\\[-2pt]
{}\\${ Fig. 5: Relative nuclear contribution $s$ as a function of
the angle of the detector for  $\overline {\,\theta
^2_{z}}_{1}=0.23\cdot10^{-4}$ cm$^{-1}$ (solid curve), $\overline
{\,\theta ^2_{z}}_{2}=0.23\cdot10^{-3}$ cm$^{-1}$ (dashed curve)
and $\overline {\,\theta ^2_{z}}_{3}=0.23\cdot10^{-2}$ cm$^{-1}$
(dotted curve).}

\bigskip

In the case of a thick target $\theta_{z}\gg \theta_{n}$, the
relation ${\varphi _{nn} } \mathord{\left/ {\vphantom {{\varphi
_{nn} } {\varphi _0 }}} \right. \kern-\nulldelimiterspace}
{\varphi _0 }$ at large thickness becomes independent of
$\vartheta _{\det } $ and equals its maximum value.

$$ \left( {\frac{\varphi _{nn} }{\varphi _0 }}
\right)^{\scriptscriptstyle \mathrm{I}} = \frac{1}{kR_d^2 }\left(
{\mathrm{Re}d\frac{\mathrm{Im}d_1 }{\mathrm{Re}d_1 } -
\mathrm{Im}d} \right).\eqno(4.43)
$$

The numerical value of  $\left| {{\varphi _{nn} } \mathord{\left/
{\vphantom {{\varphi _{nn} } {\varphi _0 }}} \right.
\kern-\nulldelimiterspace} {\varphi _0 }} \right|_{\max } = 0.24$.

For $\theta_{z}\ll \theta_{n}$, the contribution of  ${\varphi
_{nn} } \mathord{\left/ {\vphantom {{\varphi _{nn} } {\varphi _0
}}} \right. \kern-\nulldelimiterspace} {\varphi _0 }$ depends on
$\vartheta _{\det } $ (see Fig.5); for $\vartheta _{\det } \ll
\theta_{z}$ this relation is $\displaystyle\left( {\frac{\varphi
_{nn} }{\varphi _0 }} \right)^{\scriptscriptstyle \mathrm{I}} =
\frac{1}{kR_d^2 }\left( {\mathrm{Re}d\frac{\mathrm{Im}d_1
}{\mathrm{Re}d_1 } - \mathrm{Im}d} \right)\frac{k^2R_d^2
}{2}\overline {\,\theta ^2_{z}}$. It increases gradually  with
growing angular width of the detector and achieves the maximum
absolute value (4.43) at $\vartheta _{\det } \gg \theta_{n}$.

From this follows that the nuclear contribution $\varphi _{nn} $
can be neglected for any  $\vartheta _{\det } $. In this case the
magnitude of the relation $\vert \varphi _{nn} / {\varphi }'_0
\vert $ is of the order of  $10^{ - 2}$.

\section*{Conclusion}

Experimental observation of the phenomenon of birefringence of
particles naturally always implies such experimental arrangement,
in which the beam of particles is directed to the target, passes
through it and then the detector registers the polarization
characteristics of particles moving in the direction of incidence
of the initial beam and detected within a certain small angular
interval relative to this direction.  A comprehensive description
of the interaction between particles (deuterons) and the nuclei of
matter is given by a kinetic equation for the density matrix.
Using this equation one can also see that the scattering process
may be both coherent and incoherent.
The birefringence effect itself, and, in particular, one of its
parameters, the rotation angle of the polarization vector is the
result of a coherent interaction between the particle and the
target.
Alongside with coherent interaction, incoherent interaction  also
leads to the rotation of the polarization vector relative to the
chosen direction. That is why the total effective rotation angle
is  the characteristic, which is  measured in the experiment.
As it has been shown,  real experimental conditions also influence
the value of $\varphi_{eff}$: being defined as an integral
characteristic, it depends on the value of the  angle of the
detector collimator. In the general case, the magnitude of the
Coulomb-nuclear and nuclear-nuclear contributions increases with
growing $\vartheta_{det}$. In particular, at
$\vartheta_{det}>\theta_{n}$, the correction from the
Coulomb-nuclear interference exceeds by one order of magnitude the
main contribution of $\varphi_{0}$.
Here  arises the question about the magnitude of the total
contribution of the Coulomb-nuclear interference to the rotation
angle  $\varphi_{eff}$. It has been shown that these terms are
compensated completely only when the beam particles are registered
in a $4\pi$ geometry of the experiment and the deuteron energy is
such that one can take into account only  the first Born
approximation to the Coulomb amplitude.
Therefore, in a real experimental arrangement, the compensation of
the Coulomb-nuclear terms will not be observed. The quantitative
analysis demonstrated that  for a thin target, the magnitude of
the total Coulomb-nuclear interference for
$\vartheta_{det}\gg\theta_{n}$ equals zero with high accuracy. At
the same time, as a result of multiple scattering of particles in
the target, the contribution of the Coulomb-nuclear terms becomes
dependent on one more parameter: the average squared angle of
multiple scattering.
It has been shown that if its magnitude at depth $z$  is of the
order of or larger than the magnitude of the squared diffraction
angle of nuclear scattering, the Coulomb-nuclear interference can
be neglected for any angles of the detector. If this condition is
not fulfilled,  the contribution of the total Coulomb-nuclear
interference to the rotation angle should be taken into account.

The contribution of the nuclear-nuclear interaction for the two cases of target thicknesses considered above is
small as compared to the main contribution.

Thus, the elastic coherent scattering of deuterons in the target
leads to  additional corrections to the birefringence effect, the
magnitude of these contributions appreciably depends on both the
average squared angle of multiple scattering and on the specific
geometry of the experiment, namely  on the angular width of the
collimator.

{}


\begin{thebibliography}{}

\bibitem{1} Baryshevsky V.G. // Phys. Lett., 1992. Vol. 171A. P.
431-434.

\bibitem{2} Baryshevsky V.G. // J. Phys., 1993. Vol. 19G. P. 273-282.


\bibitem{3} Baryshevsky V., Dueweke C., Emmerich R. \emph{et al}. // LANL
e-print archive: hep-ex/0501045.

\bibitem{4} Seyfarth H., Engels R., Rathmann F. \emph{et al}. // Phys.
Rev. Lett., 2010. Vol. 104. 222501.doi: 10.1103.

\bibitem{5} Azhgirey L.S., Vasiliev T.A., Gurchin Yu.V. \emph{et al}. //
Physics of Particles and Nuclei Letters., 2010. Vol. 7. No 1. P.
27-32.

\bibitem{6} Baryshevsky V.G. // J. Phys.,  2008. Vol. 35G.  P.035102

\bibitem{7} Baryshevsky V. G. // Proceedings of the 6th International
Conference on Nuclear Physics at Storage Rings STORI'05, Schriften
des Forschungszentrums Julich, Matter and Materials, Vol. 30, p.
277 (2005).

\bibitem{8}  Baryshevsky V.G.,  Shyrvel  A.R. //  LANL e-print archive:
hep-ph/0503214v2.

\bibitem{9} Baryshevsky V.G., Rouba A. A. // LANL e-print archive:
hep-ph/0706.3808v2; Vesti National Akad.Nauk. of Belarus Ser.
fiz.-math. nauk, 2009. N 1. C. 64-69.

\bibitem{10} Baryshevsky V.G., Shekhtman A.G. // Phys. Rev. 1996. Vol.
53C. P. 267-276.

\bibitem{11} Baryshevsky V.G., Batrakov K.G., Cherkas S.L. // Proceedings
of International  Workshop on Quantum Systems: Quantum System '96.
Minsk, 1996. P. 142-146.

\bibitem{12} Baryshevsky V.G., Batrakov K.G., Cherkas S.L. // LANL e-print
archive: hep-ph/9907464.


\bibitem{13}  Balling L.,  Hanson R.,  Pipkin F. //  Phys. Rev.,  1964.
Vol. 133, n. 3A, P. A607–A626.

\bibitem{14}  Luttinger J.,  Kohn W. // Phys. Rev., 1958. Vol. 109 P.
1892.

\bibitem{15}  Pitaevskii L.P., Lifshitz E.M. // Physical Kinetics. in L.D.
Landau, E.M. Lifshitz // Landau L.D.,  Lifshitz E.M. //Course of
Theoretical Physics Vol. 10 (1st ed.) (Pergamon Press, 1981).


\bibitem{16} M.L. Goldberger and R.M. Watson // Collision Theory. (Wiley,
New York, 1984).

\bibitem{17} Berestetskii V.B., Lifshitz E.M., Pitaevskii L.P. // Quantum
electrodynamic // L.D. Landau, E.M. Lifshitz // Course of
Theoretical Physics Vol. 4, (2nd ed.) (Butterworth-Heinemann,
1982).

\bibitem{18} Davydov A. S. // Quantum Mechanics. (BHV-Petersburg, 2010) [in Russian] %496-581.

\bibitem{19} Baryshevsky V.G. // LANL e-print archive: hep-ph/0708.4174v1.

\bibitem{20} Czyz W., Maximon L.C. // Ann. of Physics, 1969. Vol. 52. P.
59-121.

\bibitem{21} Rouba A.A., Private communication.

\bibitem{22} Galitsky, V.M., Karnakov B.M., Kogan V.I. // Zadachi po
Kvantowoi Mekhanike [in Russian (Problems in Quantum Mechanics)]
(Nauka, Moscow, 1992)  P. 678-680.

\bibitem{23}   Ter-Mikaelian M.L. // High-Energy Electromagnetic Processes in Condensed Media (Wiley, New York, 1972).%Ter-Mikaelian  M. L. // Influence of the Medium on Electromagnetic Processes at High Energies Armenian Academy of Sciences, Erevan, 1969) P. 153-160.

\bibitem{24} Bethe H.A. // Phys. Rev., 1952. Vol. 89. P. 1256-1270.

\end{thebibliography}
\end{document}